\documentclass[aps,prb,groupeaddress,color,supperscriptaddress,showpacs,footinbib,twocolumn,showkeys,final]{revtex4}
\usepackage{graphics,graphicx,color}
\begin{document}

\title{Quantum Coherence at Low Temperatures in Mesoscopic Systems: Effect of Disorder}

\author{Yasuhiro~Niimi$^{1,2,*}$, Yannick~Baines$^{1}$, Thibaut~Capron$^{1}$, Dominique~Mailly$^{3}$, Fang-Yuh~Lo$^{4,5}$, Andreas~D.~Wieck$^{4}$, Tristan~Meunier$^{1}$, Laurent~Saminadayar$^{1,5,6,\dagger}$ and Christopher~B\"{a}uerle$^{1,\dagger}$}
\affiliation{$^{1}$Institut N\'{e}el, CNRS and Universit\'{e} Joseph
Fourier, BP 166, 38042 Grenoble, France} 
\affiliation{$^{2}$Department of Physics, Tohoku University, 6-3 Aramaki-aza Aoba, Aoba-ku, Sendai, Miyagi 980-8578, Japan} 
\affiliation{$^{3}$Laboratoire de Photonique et Nanostructures, route de Nozay, 91460 Marcoussis, France} 
\affiliation{$^{4}$Lehrstuhl f\"{u}r Angewandte Festk\"{o}rperphysik, Ruhr-Universit\"{a}t, Universit\"atsstra{\ss}e 150, 44780 Bochum, Germany}
\affiliation{$^{5}$Department of Physics, National Taiwan Normal University, 88, Sec. 4, Ting-Chou Rd., Taipei City 11677, Taiwan}
\affiliation{$^{6}$Institut Universitaire de France, 103 Boulevard Saint-Michel, 75005 Paris, France}

\date{June 8, 2010}

\begin{abstract}
We study the disorder dependence of the phase coherence time of
quasi one-dimensional wires and two-dimensional (2D) Hall bars
fabricated from a high mobility GaAs/AlGaAs heterostructure. Using
an original ion implantation technique, we can tune the intrinsic
disorder felt by the 2D electron gas and continuously vary the
system from the semi-ballistic regime to the localized one. In the
diffusive regime, the phase coherence time follows a power law as a
function of diffusion coefficient as expected in the Fermi liquid
theory, without any sign of low temperature saturation.
Surprisingly, in the semi-ballistic regime, it becomes independent
of the diffusion coefficient. In the strongly localized regime we
find a diverging phase coherence time with decreasing temperature,
however, with a smaller exponent compared to the weakly localized
regime.
\end{abstract}


\maketitle

\section{Introduction}

Quantum coherence in mesoscopic systems is one of the major issues
in modern condensed matter physics as it is intimately linked to the
field of quantum information. The interaction of solid state qubits
with environmental degrees of freedom strongly affects the fidelity
of the qubit and leads to decoherence. Consequently, the decoherence
process limits significantly the performance of such devices and it
is often regarded as a nuisance. It is hence important to understand
the limitation to the electronic coherence not only from the
fundamental point of view but also for the realization of qubit
devices.

According to the Fermi liquid (FL) theory,~\cite{AAK_82} the phase
coherence time $\tau_{\phi}$ is limited by any inelastic scattering
events, such as electron-electron interactions, electron-phonon
interactions or spin-flip scattering of electrons from magnetic
impurities. In all cases, $\tau_{\phi}$ is expected to diverge as
the temperature goes to zero. Contrary to this expectation,
experimentally $\tau_{\phi}$ seems to saturate at very low
temperatures. Mohanty and coworkers have observed systematic low
temperature saturations of $\tau_{\phi}$ for Au
wires.~\cite{mohanty_prl_97} This experiment has triggered a
controversial debate whether the low temperature saturation of
$\tau_{\phi}$ is really \textit{intrinsic} or \textit{extrinsic}.
Golubev and Zaikin (GZ) have claimed that $\tau_{\phi}$
intrinsically saturates at zero temperature due to electron-electron
interactions in the ground state.~\cite{GZ_prl_98,GZ_prb_06} On the
other hand, this low temperature saturation of $\tau_{\phi}$ can
also be explained by various extrinsic reasons such as the presence
of dynamical two level systems,~\cite{imry_epl_99,zawa_prl_99} the
presence of a small amount of magnetic
impurities,~\cite{birge+pierre_prl_02,pierre_prb_03,schopfer_prl_03,bauerle_prl_05,birge_prl_06,mallet_prl_06,sami_physicaE_07,capron_prb_08,glazman_03,zarand_prl_04,Borda_07,rosch_prl_06,Micklitz_07,Costi_prl_08}
radio frequency assisted dephasing,~\cite{gershenson_prl_06}
\textit{etc}. However, none of those extrinsic mechanisms has been
able to rule out the possibility that there might be an intrinsic saturation
of $\tau_{\phi}$ at low temperature. For example, an extremely small
amount of magnetic impurities can always explain the observed
saturation of
$\tau_{\phi}$.~\cite{bauerle_prl_05,birge_prl_06,mallet_prl_06,sami_physicaE_07}
This fact shows that one cannot clearly discriminate the intrinsic
and extrinsic mechanisms only from the temperature dependence of
$\tau_{\phi}$ and another parameter is needed to distinguish them.

In order to settle the important debate about the decoherence at
zero temperature, we have chosen to study the disorder dependence,
in other words, the diffusion coefficient $D$ dependence of
$\tau_{\phi}$ as the two different scenarios (Fermi liquid
description or intrinsic saturation) predict different $D$
dependencies on $\tau_{\phi}$. Some attempts to measure the $D$
dependence of $\tau_{\phi}$ have been performed in metallic
systems~\cite{mohanty_prl_97,Lin_2001} as well as in semiconductor
ones.~\cite{Noguchi_JAP_96} However, any clear conclusion could not
be drawn from those experiments, since it is difficult to vary $D$
in a controlled way over a wide range.

In this article, we report on the electronic phase coherence time
$\tau_{\phi}$ measurements in quasi one-dimensional (1D) wires and
two-dimensional (2D) Hall bars fabricated from a high mobility 2D
electron gas (2DEG). Using an original ion implantation technique,
as detailed in the next section, we can vary the diffusion
coefficient $D$ over three orders of magnitude without changing any
other parameter, such as electron density, band structure
\textit{etc}. In our previous work on the low temperature
decoherence as a function of $D$,~\cite{niimi} we have presented
mainly results for one quasi-1D wire. Here we present an exhaustive
report concerning the disorder dependence for quasi-1D wires as well
as 2D Hall bars. The dimensionality defined in this paper is
determined in terms of the phase coherence length
$L_{\phi}=\sqrt{D\tau_{\phi}}$ as follows; when $L_{\phi}$ is larger
than the width of wire $w$ but smaller than the length of wire $L$,
the system is ``quasi-1D". On the other hand, when $L_{\phi} \ll w <
L$, it is ``2D". Depending on the range of the diffusion coefficient
$D$, several different regimes can be attained for quasi-1D systems,
i.e. ballistic, semi-ballistic, diffusive, and strongly localized
regimes. In this work, we present decoherence measurements in the
semi-ballistic, diffusive, and strongly localized regimes for the
quasi-1D system as well as in the weakly and strongly localized
regimes for the 2D system.

The article is organized as follows; in the next section,
experimental details are described. In Sec. III, we review theories
on the phase coherence time and weak localization (WL) in the
diffusive (or weakly localized) regime, and then present
experimental results in this regime. The results on the WL curves
and the phase coherence time in the semi-ballistic regime are
presented in Sec. IV. Section V is devoted to the discussion of the
disorder dependence of the decoherence in the quasi-1D wires. In
Sec. VI, we discuss the \textit{effective electron temperature} in
our samples as it is a very important issue when discussing
decoherence at zero temperature. Finally, in Sec. VII we present
data for decoherence in the strongly localized regime.

\section{Sample fabrication and experimental set-up}

Samples have been fabricated from a GaAs/AlGaAs heterostructure
grown in ultra high vacuum by molecular beam epitaxy with electron
density $n_{e} = 1.76 \times 10^{11}$ cm$^{-2}$ and mobility
$\mu_{e} = 1.26 \times 10^{6}$ cm$^{2}/$V$\cdot$s at a temperature
of $T=4.2$ K in the dark and before processing. All lithographic
steps are performed using electron beam lithography on
polymethyl-methacrylate (PMMA) resist. Firstly, ohmic contacts have
been patterned by evaporating an AuGeNi alloy onto the wafer. The
wafer has been subsequently annealed at 450 $^{\circ}$C for a few
minutes in a hydrogen atmosphere. Secondly, our desired
nanostructures (wires, Hall bars, \textit{etc}) have been etched
into the MESA by argon ion milling over a depth of 5 nm using an
aluminium mask. The mask has then been removed with a NaOH solution.
Such a \textit{shallow etching} results in highly specular
reflection on the boundaries of the sample,~\cite{shallow_etching}
as discussed in Sec. IV B.

\begin{figure}
\begin{center}
\includegraphics[width=6cm]{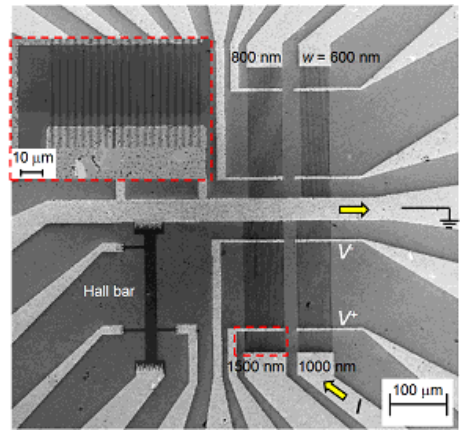}
\caption{(Color online) Scanning Electron Microscopy (SEM) image of the sample. The dark and white parts represent the mesas and electrodes,
respectively. The voltage probes for the 1000 nm wide wires as well as the ground and current bias are added in the figure.} \label{sem_image}
\end{center}
\end{figure}

A scanning electron micrograph (SEM) of a typical sample used in
this work is shown in Fig.~\ref{sem_image}. Each sample consists of
4 sets of wires of length $L=150$ $\mu$m and of lithographic width
$w =$ 600, 800, 1000 and 1500 nm. In order to suppress universal
conductance fluctuations (UCFs), each set consists of 20 wires
connected in parallel. In addition, a Hall bar allows to measure the
electronic parameters of the 2DEG: $n_{e}$, $\mu_{e}$, elastic mean
free path $l_{e}$, elastic scattering time $\tau_{e}$, \textit{etc}.
The diffusion coefficient is obtained via the relation $D = 1/2
(v_{F}l_{e})$ where $v_{F}$ is the Fermi velocity. We summarize the
fomulas for the electronic parameters in Table~\ref{table1}.

\begin{table}[bmp]
\renewcommand{\thefootnote}{\fnsymbol{footnotemark}}
\caption{Formulas of some electronic parameters. The Drude conductivity $\sigma=\frac{1}{R_{xx}}\frac{L}{w}$ is obtained from the Hall bar.}
\label{table1}
\begin{ruledtabular}
\begin{tabular}{ll}
Electron density $n_{e}$& $\displaystyle n_{e}=\frac{B}{eR_{xy}}$ or $\displaystyle n_{e}=\frac{eB\nu}{h}$\footnotemark[0]{$^\dag$} \\
Fermi velocity $v_{F}$ & $\displaystyle v_{F}=\frac{\hbar k_{F}}{m^{*}}=\frac{\hbar\sqrt{2\pi n_{e}}}{m^{*}}$ \\
Elastic scattering time $\tau_{e}$ & $\displaystyle \tau_{e}=\frac{m^{*} \sigma}{e^{2}n_{e}}$ \\
Elastic mean free path $l_{e}$ & $\displaystyle l_{e}=v_{F}\tau_{e}=\frac{h \sigma}{e^{2}\sqrt{2\pi n_{e}}}$ \\
Diffusion coefficient $D$& $\displaystyle D=\frac{1}{2}v_{F}l_{e}=\frac{\pi \hbar^{2}\sigma}{e^{2}m^{*}}$ \\
Electron mobility $\mu_{e}$ & $\displaystyle \mu_{e}=\frac{e\tau_{e}}{m^{*}}=\frac{\sigma}{n_{e}e}$ \\
$k_{F}l_{e}$ & $\displaystyle k_{F}l_{e}=\frac{h}{e^{2}}\sigma$ \\
\end{tabular}
\end{ruledtabular}
\footnotetext[0]{$^{\dag} \nu$ is the filling factor.}
\end{table}

\begin{figure}
\begin{center}
\includegraphics[width=7cm]{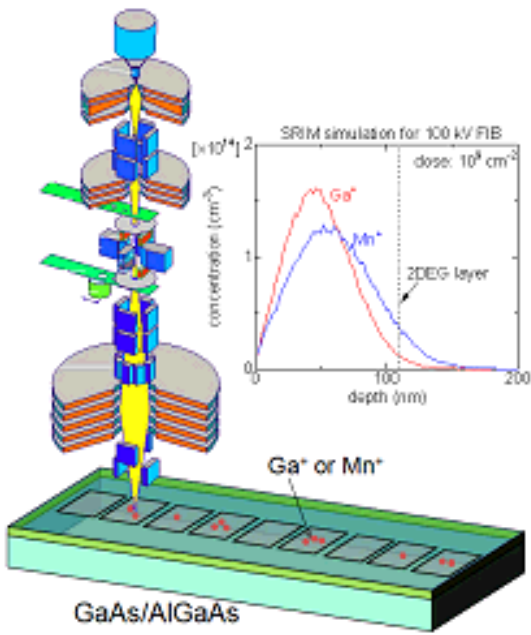}
\caption{(Color online) Schematic drawing of a FIB microscope placed
on the GaAs wafer. The inset shows an SRIM simulation 
(see Ref.~\onlinecite{fib}) of
the implanted ion concentration as a function of depth at a dose of
$10^{9}$ cm$^{-2}$ and at an energy of 100 keV. The ions are
predominantly implanted 50 nm above the 2DEG.} \label{fib}
\end{center}
\end{figure}

A large number of such samples is fabricated on the same wafer. In
order to vary the disorder in our samples, we place a Focused Ion
Beam (FIB) microscope coupled to an interferometric stage on one
sample using several alignment marks written on the wafer
[Fig.~\ref{fib}]. We then implant locally Ga$^{+}$ or Mn$^{+}$ ions
with an energy of 100 keV into the sample. For such an energy, the
implanted ions penetrate only about 50~nm into the GaAs
heterostructure,~\cite{fib} whereas the 2DEG lies 110 nm below the
surface [inset of Fig.~\ref{fib}].~\cite{note_implantation} For the
doses used here, the ions create crystal defects in the AlGaAs doped
layer and modify the electrostatic disorder potential felt by the
electrons. With this original set-up we are thus able to change the
intrinsic disorder of the samples on the same wafer by simply
changing the implantation dose.  For such low doses, the implanted
ions affect only the elastic scattering time and the mobility of the
itinerant electrons in the 2DEG,~\cite{wieck_pss_08} but do not
affect the band structure and the effective mass of
GaAs.~\cite{wieck_surf_sci_90,note_high_implantation_dose}

\begin{figure}
\begin{center}
\includegraphics[width=6cm]{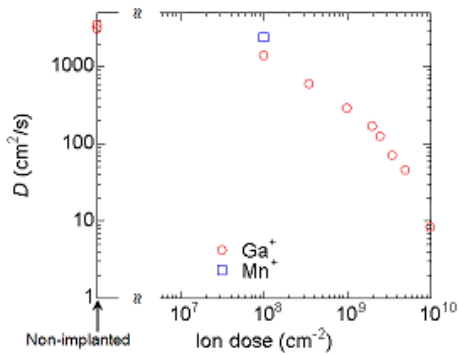}
\caption{(Color online) Diffusion coefficient as a function of ion dose for Ga$^{+}$ and Mn$^{+}$.}
\label{ion_dose}
\end{center}
\end{figure}

By varying the implantation dose for different samples from $10^{8}$
to $10^{10}$ cm$^{-2}$, we are able to vary the diffusion
coefficient from 3500 cm$^{2}$/s (unimplanted sample) to 8
cm$^{2}$/s. The diffusion coefficient variation as a function of
implantation dose is shown in Fig.~\ref{ion_dose}. Above an
implantation dose of $10^{9}$ cm$^{-2}$, we observe an important
variation of the diffusion coefficient. The electronic parameters of
all our samples are listed in Table~\ref{table2}.
These parameters have been measured at $T=1$ K for $D \geq
1400$ cm$^{2}/$s and 10 K for $D\leq 600$ cm$^{2}/$s.~\cite{note_D}

\begin{table*}
\caption{Characteristics of all our samples.}
\label{table2}
\begin{ruledtabular}
\begin{tabular}{ccccccccc}
Ga$^{+}$ ion dose& $D$ & $l_{e}$ & $\mu_{e}$ &$n_{e}$ &$v_{F}$ & $k_{F}l_{e}$ & $T^{*} \equiv \hbar/(k_{B}\tau_{e})$ & $B^{*} \equiv m^{*}/(e\tau_{e})$ \\
(cm$^{-2}$) & (cm$^{2}/$s) & (nm) & (cm$^{2}/$V$\cdot$s) & ($\times$10$^{11}$ cm$^{-2}$)& ($\times$10$^{7}$cm/s)&  & (K) & (G) \\
 \hline
0&3500&4000&6.2$\times$10$^{5}$&1.56&1.7&400 & 0.33 & 160 \\
0&3100&3600&5.5$\times$10$^{5}$&1.56&1.7&350 & 0.36 & 180 \\
1.0$\times$10$^{8}$\footnotemark[0]{*}&2400&2800&4.4$\times$10$^{5}$&1.49&1.7&270 & 0.46 & 230 \\
1.0$\times$10$^{8}$&1400&1700&2.6$\times$10$^{5}$&1.50&1.7&160 & 0.78 & 390 \\
6.0$\times$10$^{8}$&600&660&9.7$\times$10$^{4}$&1.72&1.8&69 & 2.1 & 1000 \\
1.0$\times$10$^{9}$&290&340&5.2$\times$10$^{4}$&1.52&1.7&33 & 3.9 & 1900 \\
2.0$\times$10$^{9}$&170&200&3.1$\times$10$^{4}$&1.48&1.7&19 & 6.6 & 3300 \\
2.5$\times$10$^{9}$&130&160&2.5$\times$10$^{4}$&1.43&1.7&15 & 8.3 & 4100 \\
3.5$\times$10$^{9}$&71&95&1.7$\times$10$^{4}$&1.16&1.5&8.1 & 12 & 6000 \\
5.0$\times$10$^{9}$&46&60&1.0$\times$10$^{4}$&1.23&1.5&5.3 & 19 & 9500 \\
1.0$\times$10$^{10}$&8&12&2.4$\times$10$^{3}$&0.94&1.3&0.95 & 81 & 40000 \\
\end{tabular}
\end{ruledtabular}
\footnotetext[0]{$^{*}$Mn$^{+}$ ions are implanted.}
\end{table*}

\begin{figure}
\begin{center}
\includegraphics[width=5cm]{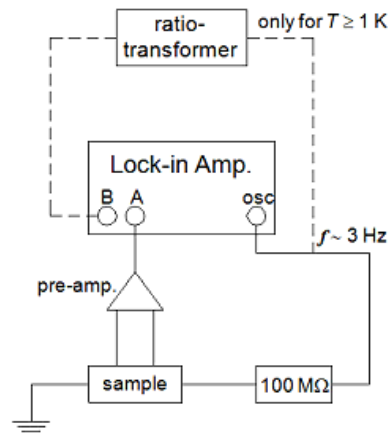}
\caption{Schematic drawing of our electric circuit. A ratio
transformer is used to subtract the background resistance and to
extract the small WL signal above 1 K.} \label{circuit}
\end{center}
\end{figure}

All measurements have been performed at temperatures down to 10 mK
using a dilution refrigerator. The resistance of the sample is
measured in a current source mode with a standard ac lock-in
technique. A voltage generated from a signal generator (typically at
a frequency of 3 Hz) is fed into the sample via a very stable
resistance, typically of the order of 10$-$100 M$\Omega$. The
voltage across the quantum wire or the Hall bar is then measured
between two voltage probes [see Fig.~\ref{sem_image}] and amplified
by a home made pre-amplifier situated at room temperature. This
voltage amplifier has an extremely low noise voltage of about 0.5
nV$/\sqrt{\rm Hz}$. Since the WL quantum correction above $\sim$1 K
is relatively small compared to classical background resistance ($<
10^{-2}$), we have used a ratio transformer in a bridge
configuration to compensate the large background signal. This allows
us to increase the sensitivity of the WL measurement. A schematic
drawing of the measuring circuit is shown in Fig.~\ref{circuit}. In
order to avoid radio-frequency heating due to external noise, all
measuring lines are extremely well filtered with commercially
available highly dissipative coaxial cables, i.e.
\textit{thermocoax}~\cite{thermocoax,JAP_Filtres} at low
temperatures and with $\pi$ filters situated at room temperature.
The total attenuation at low temperature is more than $-400$ dB at
20 GHz.  All experiments have been performed in thermal equilibrium
which means that the applied voltage across the entire sample is
kept such that the inequality $eV \leq k_{B}T$ is satisfied at all
temperatures.

\section{Diffusive regime}

\subsection{Theory}
\subsubsection{Phase coherence time}

In the weakly localized regime where $k_{F}l_{e} \gg 1$, the phase
coherence time of electrons in a conductor is limited by inelastic
scattering such as electron-electron (e-e) interactions,
electron-phonon (e-ph) interactions, the interaction with magnetic
impurities (mag), or two level systems (TLS) \textit{etc}. In the
presence of several decoherence mechanisms, the phase coherence time
$\tau_{\phi}$ can be expressed as
\begin{eqnarray}
\frac{1}{\tau_{\phi}}=\frac{1}{\tau_{\rm e-e}}+\frac{1}{\tau_{\rm e-ph}}+\frac{1}{\tau_{\rm mag}}+\frac{1}{\tau_{\rm TLS}}+\cdots. \nonumber
\end{eqnarray}
In the absence of extrinsic sources of decoherence, the phase
coherence time at low temperatures is simply dominated by e-e
interactions.~\cite{note_e_ph} Thus, hereafter, we focus on the
decoherence only due to e-e interactions.

In the FL theory without any disorder, the lifetime of
quasi-particles follows a $(E-E_F)^{-2}$ power law, with $E$ the
energy and $E_F$ the Fermi energy. In a real conductor, however,
there is disorder. Altshuler, Aronov and Khmelnitsky~(AAK) took into
account the disorder and the dimensionality of a conductor within
the framework of the FL theory.~\cite{AAK_82} AAK showed that for a
quasi-1D wire, the phase coherence time due to the e-e interactions
can be expressed by
\begin{eqnarray}
\frac{1}{\tau_{\rm e-e}^{\rm 1D}}&=&aT^{2/3} \label{AAK_1D_1}\\
&\equiv&\alpha_{\rm AAK}D^{-1/3}T^{2/3} \label{AAK_alpha}\\
&=&\frac{1}{2}\left( \frac{k_{B}\pi}{w_{\rm eff} m^{*}} \right)^{2/3}D^{-1/3}T^{2/3} \label{AAK_1D}
\end{eqnarray}
where $k_{B}$ is the Boltzmann constant and $m^{*}$ is
the effective mass of the electron.
For a 2DEG made from a GaAs/AlGaAs heterostructure, $m^{*}=0.067 m_{e}$ where
$m_{e}$ is the bare electron mass.
$w_{\rm eff} $ is the effective width of the wire which is different from the
lithographic width $w$ given in the previous section because of
lateral depletion effects inherent to the etching process.
It should be noted that Eq.~(\ref{AAK_1D})
is valid \textit{only in the diffusive regime}
where the effective width $w_{\rm eff} $ is larger than
the elastic mean free path $l_{e}$ such that the electron motion
from one boundary to the other is diffusive.

In a similar way, the phase coherence time due to the e-e
interactions for the 2D system is calculated as follows:
\begin{eqnarray}
\frac{1}{\tau_{\rm e-e}^{\rm 2D}} \simeq \frac{k_{B}T}{2m^{*}D}\ln\left( \frac{2m^{*}D}{\hbar} \right) \label{AAK_2D_1}
\end{eqnarray}
where $\hbar$ is the reduced Planck constant. Note that this
expression is valid until the thermal length $L_{T}=\sqrt{\hbar
D/k_{B} T}$ is larger than $l_{e}$. At higher temperatures such that
$L_{T} \ll l_{e}$
(or $T \gg T^{*} \equiv \hbar/(k_{B}\tau_{e})$), 
the dephasing process is not limited by disorder but simply by temperature as
expected in the FL theory without disorder:~\cite{fukuyama}
\begin{eqnarray}
\frac{1}{\tau_{\rm e-e}^{\rm 2D}} \simeq \frac{m^{*} k_{B}^{2}T^{2}}{4\hbar^{3}n_{e}}\ln\left( \frac{2 \pi \hbar^{2} n_{e}}{k_{B}T m^{*}} \right)  \label{AAK_2D_2}.
\end{eqnarray}
In semiconductors, the crossover temperature $T=\hbar/(k_{B}\tau_{e}$)
is of order of 1 K.~\cite{note_crossover_T}

\subsubsection{Weak localization correction}

The measurements of the phase coherence time can be done in various
ways such as measurements of
WL,~\cite{pierre_prb_03,sami_physicaE_07} Aharonov-Bohm conductance
oscillations,~\cite{birge+pierre_prl_02,schopfer_prl_07,texier_cmat_09}
UCFs,~\cite{birge_prb_99,mohanty_prl_03} persistent currents
\cite{sami_encyclopedia} \textit{etc}. In this work, we have chosen
to measure the phase coherence time of electrons via WL. Using this
method, one can make the most reliable and quantitative discussion
on the phase coherence time as shown in previous
works.~\cite{mohanty_prl_97,schopfer_prl_03,pierre_prb_03,bauerle_prl_05,birge_prl_06,mallet_prl_06,sami_physicaE_07,capron_prb_08,niimi}
The principle of this technique relies on constructive interference
of closed electron trajectories which are ``traveled" in opposite
direction (time reversed paths). This leads to an enhancement of the
resistance. The magnetic field $B$ destroys these constructive
interferences, leading to a negative magnetoresistance $R(B)$ (or
positive magnetoconductance $G(B)$) whose amplitude and width are
directly related to the phase coherence time.

For a quasi-1D diffusive wire where $w_{\rm eff}  > l_{e}$, the WL
correction is calculated as below:~\cite{gilles_book}
\begin{eqnarray}
\Delta G(B) &\equiv& G(B)-G(0) \nonumber\\
&=& -2N\frac{e^{2}}{h} \frac{L_{\phi}}{L} \left\{ \frac{1}{\sqrt{1+\frac{L_{\phi}^{2}w_{\rm eff} ^{2}}{3l_{B}^{4}}}} - 1 \right\} \label{HLN}
\end{eqnarray}
where ${e^{2}}/{h}$ is the quantum of conductance ($e$ is the charge
of the electron and $h$ is the Planck constant),
$l_{B}=\sqrt{\hbar/eB}$ is the magnetic length and $N$ is the number
of wires in parallel ($N=20$ in the present case). The spin-orbit
term has been neglected as spin-orbit coupling is very weak in
GaAs/AlGaAs heterostructures. As discussed later on, we can obtain
$w_{\rm eff}$ and $G(0)$ independently from the experimentally
measured magnetoconductance and therefore the \textit{only} fitting
parameter is $L_{\phi}$. By fitting the experimental
magnetoconductance $G(B)$ with Eq.~(\ref{HLN}),
we can obtain the phase coherence length
$L_{\phi}$ at any temperature. The phase coherence time
$\tau_{\phi}$ is then extracted from the relation
$L_{\phi}=\sqrt{D\tau_{\phi}}$. We note that Eq.~(\ref{HLN}) holds
only when the magnetic field satisfies the inequality $l_{B}>w_{\rm
eff} $.~\cite{Beenakker_ssp_91} When $l_{B}<w_{\rm eff} $, the
lateral confinement becomes irrelevant for the WL and a crossover
from 1D to 2D WL occurs.

If $L_{\phi} \ll w$, the 2D WL correction
to the conductance is applied and given by
\begin{eqnarray}
\Delta G(B) &=& \frac{e^{2}}{\pi h} \frac{w}{L} \left\{ \Psi \left(\frac{1}{2}+\frac{l_{B}^{2}}{4L_{\phi}^{2}} \right)  \right. \nonumber\\
&-& \left. \Psi \left(\frac{1}{2}+\frac{l_{B}^{2}}{2l_{e}^{2}} \right) + \ln \left( \frac{2L_{\phi}^{2}}{l_{e}^{2}} \right) \right\}, \label{HLN_2D}
\end{eqnarray}
where $\Psi(x)$ is the digamma function.
The digamma function has the asymptotic approximation
$\Psi(\frac{1}{2}+x) \simeq \ln x$ for large $x$.
In the case of 2D WL, the characteristic
field $B_{c}=\hbar/4eL_{\phi}^{2}$ which corresponds to one flux quantum
through an area of the order of ${L_{\phi}}^{2}$ is usually very small.
For example, if $L_{\phi}=1$ $\mu$m, $B_{c}=1.6$ G.
The suppression of the WL effect is complete
when $B > \hbar/2el_{e}^{2}$. These fields are always much
weaker than classically strong fields
$B^{*} \equiv m^{*}/(e\tau_{e})$.

\subsection{Experimental results}
\subsubsection{Quasi-1D wires}

\begin{figure}
\begin{center}
\includegraphics[width=6cm]{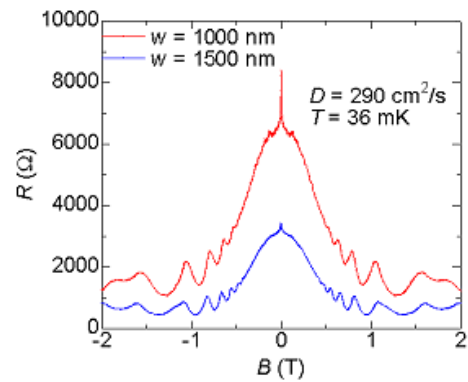}
\caption{(Color online) Magnetoresistance curves of 1000 and 1500 nm wide wires at $T=36$ mK and $D=290$ cm$^{2}$/s.}
\label{SdH_3ppm_30mK}
\end{center}
\end{figure}

In order to determine the phase coherence length $L_{\phi}$, we have
performed standard magnetoresistance measurements as a function of
temperature. A typical example for such a magnetoresistance curve is
displayed in Fig.~\ref{SdH_3ppm_30mK}. Let us first concentrate on
the field range up to a magnetic field of 2\,T. A sharp peak which
is due to WL is clearly seen at zero field. With increasing the
magnetic field the WL peak disappears and another type of negative
magnetoresistance is observed which is due to magnetic focusing.
When going to even higher fields ($>$ 0.5 T) the well-known
Shubnikov de Haas (SdH) oscillations appear.

\begin{figure}
\begin{center}
\includegraphics[width=7.5cm]{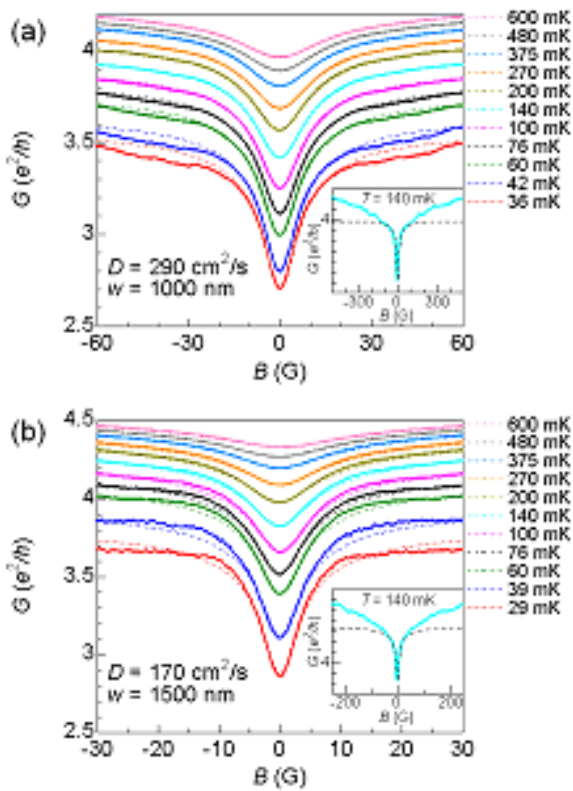}
\caption{(Color online) WL curves of (a) 1000 and (b) 1500 nm wide wires at $D=290$
and 170 cm$^{2}$/s, respectively. The conductance here is divided by
$e^{2}/h$. The broken lines are the best fits of Eq.~(\ref{HLN}).
The insets in (a) and (b) show the magnetoconductance at $T=140$ mK
in larger field ranges.} \label{WL_curves_D_290}
\end{center}
\end{figure}

Analyzing the WL peak allows to obtain the phase coherence length
$L_{\phi}$. In Fig.~\ref{WL_curves_D_290}, we show
magnetoconductance curves in units of $e^{2}/h$ for $w=1000$ and
1500 nm wide wires at different temperatures. Note that the field
scale is about three orders of magnitude smaller than that in
Fig.~\ref{SdH_3ppm_30mK}. Since we are in a diffusive regime where
$l_{e}$ is smaller than $w$, the standard WL formula Eq.~(\ref{HLN})
can be used. In Eq.~(\ref{HLN}), there are two parameters, i.e.
$L_{\phi}$ and $w_{\rm eff} $. The effective width $w_{\rm eff} $,
however, is determined by fitting the magnetoconductance at a given
temperature and diffusion coefficient. For lithographic widths
$w=1000$ and 1500 nm, we obtain $w_{\rm eff} =630$ and 1130 nm,
respectively. The effective width is then kept fixed for the entire
fitting procedure and $L_{\phi}$ remains the only fitting parameter.

The observed WL curves are nicely fitted using Eq.~(\ref{HLN}) over
the field ranges of $\pm 60$ and $\pm 30$ G for $w=1000$ and 1500
nm, respectively. At a higher field (above $\sim 100$ G), however,
the measured WL curves start to deviate from the theoretical
fittings [insets of Fig.~\ref{WL_curves_D_290}]. For this reason,
when we fit the magnetoconductance with the standard theory, we
limit the field scale within $l_{B}>w_{\rm eff} $, i.e. $|B|<15$ and
5 G for $w_{\rm eff} =630$ and 1130 nm, respectively.

\begin{figure}
\begin{center}
\includegraphics[width=6cm]{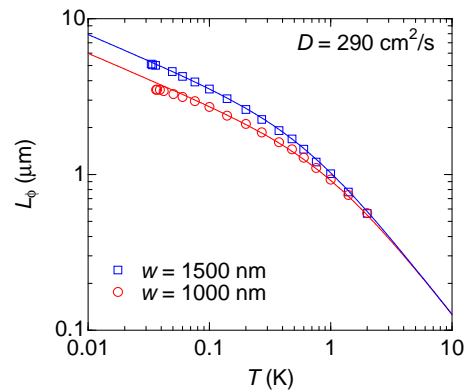}
\caption{(Color online) Phase coherence length of 1000 and 1500 nm wide wires as a function
of $T$ at $D=290$ cm$^{2}$/s. The solid lines are the best fits
with Eq.~(\ref{AAK_fitting_1D}).}
\label{L_phi_D_290}
\end{center}
\end{figure}

The extracted phase coherence length $L_{\phi}$ is plotted as a
function of $T$ at $D=290$ cm$^{2}/$s for $w=1000$ and 1500 nm wide
wires in Fig.~\ref{L_phi_D_290}. At low temperatures, $L_{\phi}$
nicely follows a $T^{-1/3}$ law down to the lowest temperatures for
both the wires. Note that the temperature below 40 mK has been
corrected by measuring \textit{in situ} the electron temperature of
the quasi-1D wire based on e-e interaction corrections as detailed
in Sec. VI. The absolute values of $L_{\phi}$ at low temperatures
are different between the two wires, which is expected in the AAK
theory in Eq.~(\ref{AAK_1D}). 
Similar temperature dependence of
$L_{\phi}$ has also been observed in GaAs/GaAlAs
networks.~\cite{ferrier_PRL_04}

Above $\approx$ 1 K, $L_{\phi}$
follows a $T^{-1}$ law and its absolute value does not depend on the
width of the wire. This is because $L_{\phi}$ is not limited by
disorder any more but follows the FL theory without disorder as
shown in Eq.~(\ref{AAK_2D_2}).~\cite{Noguchi_JAP_96,fukuyama} When
we fit the $L_{\phi}$ vs $T$ curves, the following equation is used:
\begin{eqnarray}
L_{\phi} = \sqrt{D\tau_{\phi}} = \sqrt{\frac{D}{a_{\rm exp}T^{2/3}+b_{\rm exp}T^{2}}}, \label{AAK_fitting_1D}
\end{eqnarray}
where $a_{\rm exp}$ and $b_{\rm exp}$ are the fitting parameters.~\cite{note_fitting_L_phi}

\subsubsection{Hall bars}
\begin{figure}
\begin{center}
\includegraphics[width=8cm]{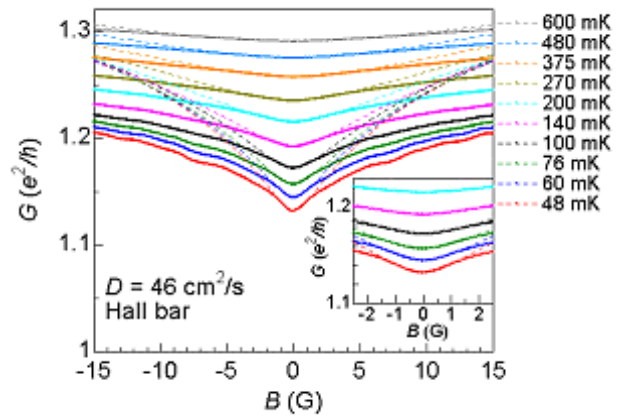}
\caption{(Color online) Magnetoconductance curves of a Hall bar at $D=46$ cm$^{2}$/s at different temperatures. The conductance is normalized by $e^{2}/h$. The broken lines are the best fits to Eq.~(\ref{HLN_2D}). The fitted curves deviate from the experimental data at around $B_{c}$. The inset shows a closeup view of the low field part of the magnetoconductance at low temperatures.}
\label{WL_curves_D_46}
\end{center}
\end{figure}

In a similar manner to the quasi-1D case, the phase coherence length
for Hall bars can also be extracted
by fitting the WL curves
with Eq.~(\ref{HLN_2D}).~\cite{Renard_prb_2005,2DEG_prb_2006}
Figure~\ref{WL_curves_D_46} shows the WL curves of the Hall bar
at $D=46$ cm$^{2}$/s
at different temperatures and the best fits with Eq.~(\ref{HLN_2D}).
For these fittings we restrict the field scale
to $B_{c}=\hbar/4eL_{\phi}^{2}$ for which
2D WL formula is applicable.~\cite{2DEG_prb_2006,2DHG_prb_2004}
We recall that $B_{c}$ is of the order of 1 G
when $L_{\phi}=1$ $\mu$m [see inset of Fig.~\ref{WL_curves_D_46}].
With increasing temperature, $L_{\phi}$ becomes smaller
and the fitting region becomes larger as shown in Fig.~\ref{WL_curves_D_46}.
This clearly justifies the field limitation for the fittings.

\begin{figure}
\begin{center}
\includegraphics[width=6cm]{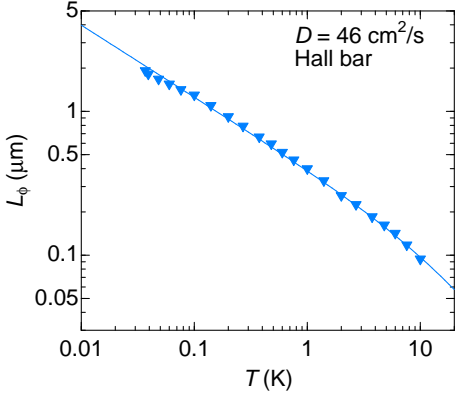}
\caption{(Color online) Phase coherence length of the Hall bar at $D=46$ cm$^{2}$/s as a function of $T$. The solid line is the best fit with Eq.~(\ref{AAK_fitting_2D}).}
\label{L_phi_D_46}
\end{center}
\end{figure}

The obtained $L_{\phi}$ of the Hall bar is plotted as a function of
$T$ in Fig.~\ref{L_phi_D_46}. At low temperatures, it follows a
$T^{-1/2}$ law as expected in the AAK theory for 2D systems [see
Eq.~(\ref{AAK_2D_1})]. On the other hand, $L_{\phi}$ has a $T^{-1}$
dependence above $\approx$ 5 K where the thermal length $L_{T}$ is
smaller than $l_{e}$.~\cite{fukuyama} The whole $L_{\phi}$ vs $T$
curve of the Hall bar is fitted by combining Eqs.~(\ref{AAK_2D_1})
and (\ref{AAK_2D_2}) as below:
\begin{eqnarray}
L_{\phi} = \sqrt{D\tau_{\phi}} = \sqrt{\frac{D}{a_{\rm exp}T+b_{\rm exp}T^{2}}}, \label{AAK_fitting_2D}
\end{eqnarray}
where $a_{\rm exp}$ and $b_{\rm exp}$ are the fitting parameters.
The $\ln(T)$ term in Eq.~(\ref{AAK_2D_2}) has been neglected here as we only
measure the low temperature regime.

\section{Semi-ballistic regime}

\subsection{Theory}
In this subsection, we review the WL theory for quasi-1D wires
in the semi-ballistic regime where $w_{\rm eff}  < l_{e} \ll L$.
The WL in this regime has been studied theoretically
by Beenakker and van Houten (BvH).~\cite{Beenakker_prb_88}
In such a clean limit, it is necessary to take into account specular
reflections on the boundary of the wires and flux cancelation
effects. Especially, the flux cancellation effect is of importance
in the pure conductor regime, where the electrons move ballistically
from one wall to the other. This effect leads to a wider WL curve
compared to the diffusive case.

The WL correction in the semi-ballistic regime has been calculated
by modifying the standard WL formula
Eq.~(\ref{HLN}):~\cite{Beenakker_prb_88}
\begin{eqnarray}
\Delta G(B)=-2N\frac{e^{2}}{h}\frac{L_{\phi}}{L}
\left\{ \left( \frac{1}{\sqrt{1+\frac{L_{\phi}^{2}}{D\tau_{B}}}} - 1 \right) \right. \nonumber\\
- \left. \left(\frac{1}{\sqrt{1+\frac{L_{\phi}^{2}}{D\tau_{B}}+\frac{2L_{\phi}^{2}}{l_{e}^{2}}}} -\frac{1}{\sqrt{1+\frac{2L_{\phi}^{2}}{l_{e}^{2}}}} \right)
\right\}, \label{eq_bvh}
\end{eqnarray}
where $\tau_{B}$ is the magnetic scattering time. The first two
terms are the same as Eq.~(\ref{HLN}) except $D\tau_{B}$ which is
different from the diffusive case as discussed below. The last two
terms come from a short-time cutoff. On short time scales
$t<\tau_{e}$, the motion is ballistic rather than diffusive, and the
return probability is expected to go to zero smoothly as one enters
the ballistic regime. The short-time cutoff, on the other hand,
should become irrelevant for $\tau_{\phi} \gg \tau_{e}$. Such a
short-time cutoff has been inserted heuristically to compensate the
ballistic motion in the WL correction.

In the semi-ballistic regime, $\tau_{B}$ has two limiting
expressions depending on the ratio $w_{\rm eff} l_{e}/l_{B}^{2}$ as
given below:~\cite{Beenakker_prb_88}
\begin{eqnarray*}
D\tau_{B} = \left\{
\begin{array}{@{\,}ll}
D\tau_{B}^{\rm low}= \frac{9.5}{2}\frac{l_{B}^{4}l_{e}}{w_{\rm eff} ^{3}}
& \mbox{for $\sqrt{w_{\rm eff} l_{e}} \ll l_{B}$} \\
D\tau_{B}^{\rm high}= \frac{4.8}{2}\frac{l_{B}^{2}l_{e}^{2}}{w_{\rm eff} ^{2}}
& \mbox{for $\sqrt{w_{\rm eff} l_{e}} \gg l_{B} \gg w_{\rm eff} $}.
\end{array}
\right.
\end{eqnarray*}
The crossover from the \textquotedblleft low\textquotedblright~field
and \textquotedblleft high\textquotedblright~field regions is well
described by the interpolation formula:
\begin{eqnarray}
D\tau_{B} &=& D\tau_{B}^{\rm low}+D\tau_{B}^{\rm high}\nonumber \\
&=& \frac{9.5}{2}\frac{l_{B}^{4}l_{e}}{w_{\rm eff} ^{3}} + \frac{4.8}{2}\frac{l_{B}^{2}l_{e}^{2}}{w_{\rm eff} ^{2}}.
\label{eq_tau_B}
\end{eqnarray}
This expression agrees well with numerical
calculations~\cite{Beenakker_prb_88} and is useful
for comparison with experiments.
The magnetic scattering time $\tau_{B}$ in Eq.~(\ref{eq_bvh}) is then replaced
by Eq.~(\ref{eq_tau_B}) within the field scale $l_{B} \gg w_{\rm eff} $.

It should be stressed, on the other hand, that
there is little knowledge on the decoherence time
in the semi-ballistic regime, unlike
the diffusive case discussed in Sec. III A.

\subsection{Experimental results}

\begin{figure}
\begin{center}
\includegraphics[width=6cm]{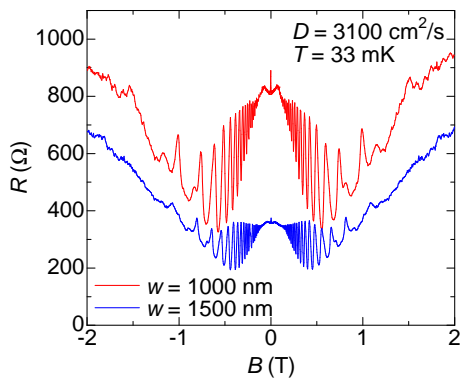}
\caption{(Color online) Magnetoresistance curves of 1000 and 1500 nm wide wires at $T=33$ mK
and $D=3100$ cm$^{2}$/s.}
\label{SdH_0ppm_20mK}
\end{center}
\end{figure}

\begin{figure}
\begin{center}
\includegraphics[width=6cm]{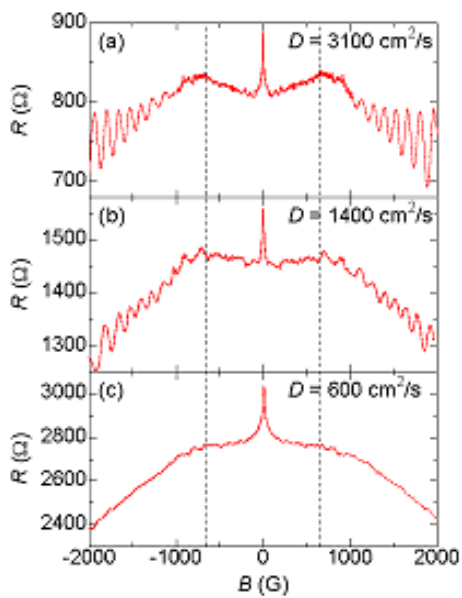}
\caption{(Color online) Magnetoresistance curves of 1000 nm wide wires at three different
diffusion coefficients; (a) $D=3100$, (b) 1400 and (c) 600 cm$^{2}$/s.
The broken line shows the maximum position of the small bump, i.e. $B_{\rm max}$.}
\label{boundary_roughness}
\end{center}
\end{figure}

As in the case of the diffusive regime, the phase coherence length
$L_{\phi}$ in the semi-ballistic regime can be extracted by fitting
experimental WL curves with Eq.~(\ref{eq_bvh}). Before discussing
the WL peak in a small field range, we show typical
magnetoresistance curves of quasi-1D wires in the semi-ballistic
regime in a field range of 2 T in Fig.~\ref{SdH_0ppm_20mK}. The
overall structure of the magnetoresistance is similar to that in the
diffusive regime [see Fig.~\ref{SdH_3ppm_30mK}]; the WL peak near
zero field and the SdH oscillation at high fields. In between these
two structures, there is a small bump due to boundary roughness
scattering~\cite{boundary_prl_89,ando_prb_91} which does not exist
in the diffusive regime. In the semi-ballistic regime where $l_{e} >
w_{\rm eff}$, the characteristics of the boundaries are of
importance. Electrons are reflected specularly on the boundary with
a given probability $p$. Otherwise they are diffusively scattered
into a random direction. In the case of shallow etching like in our
case [see also Sec. II], the specular reflection probability $p$ is
more than 80\% as reported in previous transport measurements on
2DEG samples.~\cite{shallow_etching} The diffuse boundary scattering
with a small probability $1-p$ ($<20\%$) causes the observed small
bump of the resistance in Fig.~\ref{SdH_0ppm_20mK}. In the presence
of magnetic field, the electrons follow a curved trajectory and are
scattered diffusively at each collision with the boundary. When the
cyclotron radius $R_{c}$ becomes comparable to the width of wire
($w_{\rm eff} /R_{c} \approx 0.55$),~\cite{boundary_calculation_66}
the resistance exhibits a maximum and then decreases again with
increasing field because of the absence of backscattering. As is
shown in Fig.~\ref{boundary_roughness}(a), the maximum of the bump
is located at 650 G, which corresponds to $B_{\rm max}= 0.64\hbar
k_{F}/ew_{\rm eff}$ (i.e. $w_{\rm eff} /R_{c} = 0.64$). On the other
hand, the amplitude of the bump is less than 5\% compared to the
background resistance. This result indicates that the probability of
the diffusive boundary scattering is quite
low,~\cite{boundary_prl_89} which is consistent with the above
statement (i.e. $1-p<20$\%). The observed bump structure vanishes
with decreasing $D$ or increasing disorder
[Figs.~\ref{boundary_roughness}(b) and \ref{boundary_roughness}(c)].

\begin{figure}
\begin{center}
\includegraphics[width=7cm]{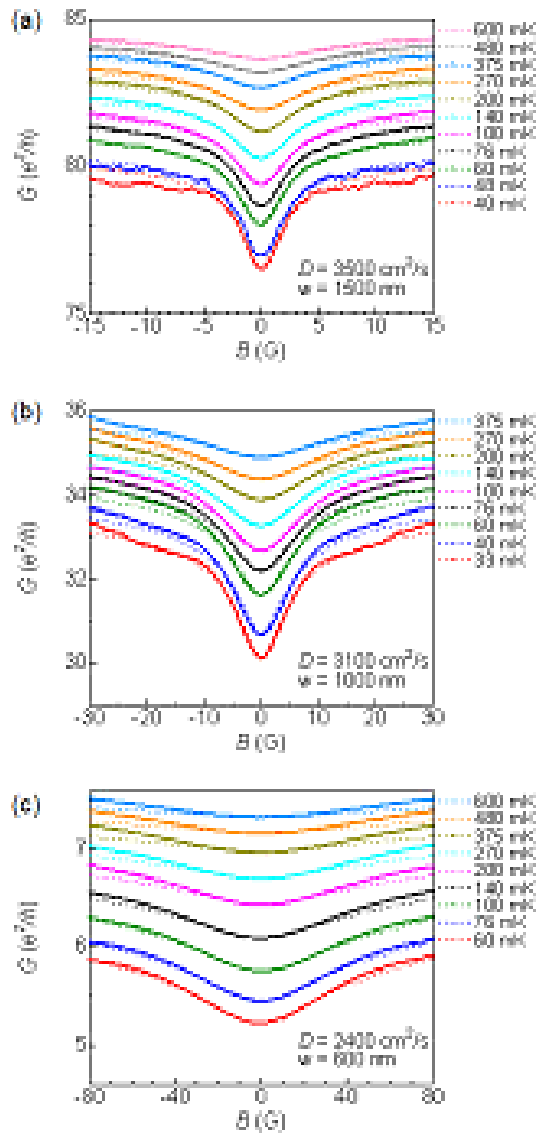}
\caption{(Color online) WL curves of (a) 1500, (b) 1000, (c) 600 nm wide wires at
$D=3500$, 3100, and 2400 cm$^{2}$/s, respectively. The conductance
is normalized by $e^{2}/h$. The broken lines are the best fits to
Eq.~(\ref{eq_bvh}).} \label{WL_curves_ballistic_regime}
\end{center}
\end{figure}

Next, we focus on the WL peak on a smaller field scale. We show
magnetoconductance curves in Fig.~\ref{WL_curves_ballistic_regime}
for three different wire widths at different temperatures. As
discussed in Sec. III B, the WL peak grows and becomes sharper with
decreasing temperature for all the wires. The width of the WL peak,
however, is almost the same as in the diffusive case [see
Fig.~\ref{WL_curves_D_290}]. This is due to flux cancelation effects
as mentioned above.

The phase coherence length $L_{\phi}$ in the semi-ballistic regime
is obtained by fitting the WL curve with Eq.~(\ref{eq_bvh}). Note
that there are three parameters in Eq.~(\ref{eq_bvh}), namely
$L_{\phi}$, $w_{\rm eff} $ and $l_{e}$. The effective width $w_{\rm
eff} $ is, however, determined in the same way as in the diffusive
case. For lithographic widths $w=1500$, 1000 and 600 nm, we obtain
$w_{\rm eff} =1130$, 630 and 230 nm,
respectively. The elastic mean free path
$l_{e}$ is also obtained from an independent measurement on the Hall
bar having the same diffusion coefficient. Thus, there is again only
one fitting parameter left, i.e. $L_{\phi}$.

\begin{figure}
\begin{center}
\includegraphics[width=6cm]{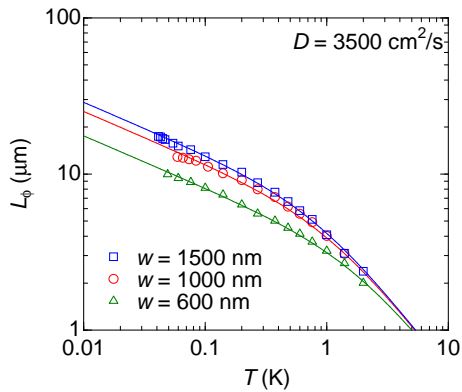}
\caption{(Color online) Phase coherence length of 1500, 1000 and 600 nm wide wires as a function of $T$ at $D=3500$ cm$^{2}$/s. The solid lines are the best fits with Eq.~(\ref{AAK_fitting_1D}).}
\label{L_phi_D_3500}
\end{center}
\end{figure}

The broken lines in Fig.~\ref{WL_curves_ballistic_regime} show the
best fits of Eq.~(\ref{eq_bvh}). The WL curves of the three wires
are nicely fitted by Eq.~(\ref{eq_bvh}) at low fields, while
deviations from the theoretical fits occur at higher fields. As
shown in the previous subsection, the BvH expression is valid only
within $l_{B} \gg w_{\rm eff}$. Therefore, for fitting the
magnetoconductance curves at any temperature we take into account
only the low field data and restrict the field range within $|B|<5$,
10 and 30 G for $w_{\rm eff} =1130$, 630 and 230 nm,
respectively.~\cite{bouchiat} Note that these fields are much larger
than $B=\hbar/ew_{\rm eff}l_{e}$ ($\sqrt{w_{\rm eff}l_{e}} \gg
l_{B}$). This means that we still have to take into account both the
``low" and ``high" field regions as pointed out in
Eq.~(\ref{eq_tau_B}). The obtained $L_{\phi}$ at $D=3500$ cm$^{2}$/s
is plotted as a function of $T$ in Fig.~\ref{L_phi_D_3500}. As in
the diffusive regime, $L_{\phi}$ follows a $T^{-1/3}$ law at low
temperatures and varies linearly with $T$ above $\approx 1$ K. Such
a temperature dependence is indeed expected in the semi ballistic
regime.~\cite{aleiner_prb_02}

\section{Disorder effect on phase coherence}

\subsection{Experimental results on quasi-1D wires}

\begin{figure}
\begin{center}
\includegraphics[width=5.5cm]{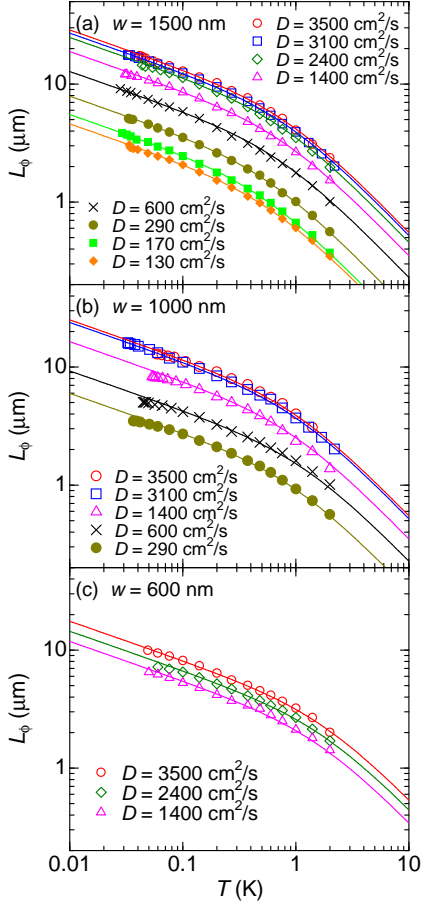}
\caption{(Color online) Phase coherence length $L_{\phi}$ of three different wires
as a function of $T$ at several different diffusion coefficients.
The open and solid symbols correspond to the semi-ballistic regime and the
diffusive one, respectively. The solid lines are the best fits to
Eq.~(\ref{AAK_fitting_1D})} \label{L_phi_all}
\end{center}
\end{figure}

\begin{figure}
\begin{center}
\includegraphics[width=6cm]{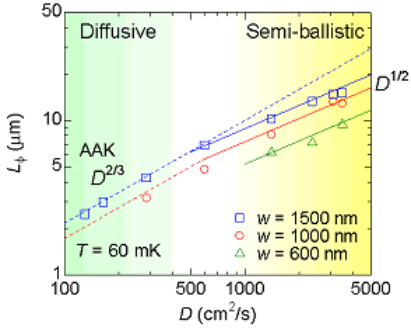}
\caption{(Color online) The diffusion coefficient dependence of the phase coherence length $L_{\phi}$ at $T=60$ mK of the three different wires. The broken and solid lines show $D^{2/3}$ and $D^{1/2}$ laws, respectively.}
\label{L_phi_vs_D}
\end{center}
\end{figure}

In Secs. III and IV, we have been discussing the temperature
dependence of the decoherence in the diffusive as well as the
semi-ballistic regimes. In this section, we will discuss the
disorder dependence of the decoherence time. For this purpose, we
first present in Fig.~\ref{L_phi_all} the temperature dependence of
the phase coherence length $L_{\phi}$ for three different wire
widths and for all investigated diffusion coefficients.
Interestingly, the temperature dependence in the low temperature
regime is identical for the diffusive regime and semi-ballistic
regime. Inspecting Fig.~\ref{L_phi_all} more closely, it is clear
that the phase coherence length $L_{\phi}$ in the semi-ballistic
regime depends more weakly on $D$ compared to the diffusive regime.
This can be emphasized by plotting the value of $L_{\phi}$ as a
function of $D$ at fixed temperature (we take $T = 60$ mK) as shown
in Fig.~\ref{L_phi_vs_D}. One clearly observes two different
$D$-dependencies. In the diffusive regime ($w_{\rm eff}
l_{e}$), $L_{\phi}$ follows a $D^{2/3}$ law, which is consistent
with the ``standard" model of decoherence proposed in the AAK
theory~\cite{AAK_82} [see Eq.~(\ref{AAK_1D})]. On the other hand, in
the semi-ballistic regime where $w_{\rm eff} < l_{e}$, $L_{\phi}$
has a different power law as a function of the diffusion
coefficient, $D^{\lambda}$ with a parameter $\lambda$ close to 1/2.
This behavior can be seen for the three different widths of the
wires. The crossover between the two regimes occurs when $w_{\rm
eff}$ becomes comparable to $l_{e}$, i.e. $D \sim 1000$ cm$^{2}/$s.

\begin{figure}
\begin{center}
\includegraphics[width=6cm]{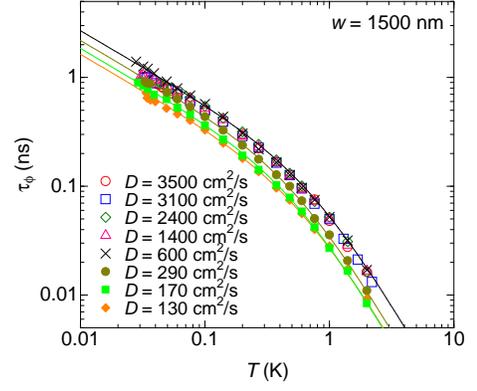}
\caption{(Color online) Phase coherence time $\tau_{\phi}$ of 1500 nm wide wires as a function of $T$ at different diffusion coefficients.}
\label{tau_phi_1500nm}
\end{center}
\end{figure}

To compare our experimental results directly with theoretical
expressions, it is more convenient to plot the diffusion coefficient
dependence of  $\tau_{\phi}$ rather than
$L_{\phi}$.~\cite{AAK_82,GZ_prl_98,GZ_prb_06} We thus obtain the
phase coherence time $\tau_{\phi}$ assuming that the relation
$L_{\phi} =\sqrt{D\tau_{\phi}}$ holds for all the investigated
diffusion coefficients. In Fig.~\ref{tau_phi_1500nm}, we show the
temperature dependence of the phase coherence time $\tau_{\phi}$ of
the 1500 nm wide wires at different $D$. At low temperatures, it
follows a $T^{-2/3}$ power law at any diffusion coefficient as
expected for the quasi-1D diffusive regime [see
Eq.~(\ref{AAK_1D_1})]. Above 1 K, $\tau_{\phi}$ tends towards a
$T^{-2}$ dependency, in accordance with the FL theory without
disorder [see Eq.~(\ref{AAK_2D_2})].

\begin{figure}
\begin{center}
\includegraphics[width=6cm]{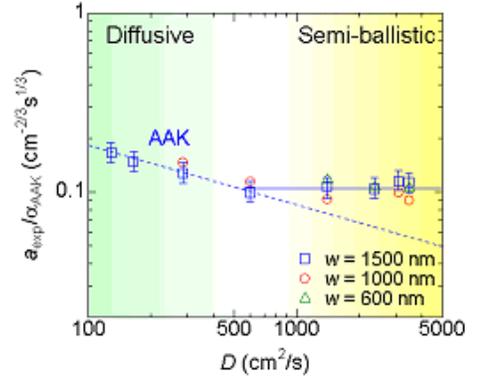}
\caption{(Color online) The experimental coefficient $a_{\rm exp}$ of Eq.~(\ref{AAK_fitting_1D}) scaled by $\alpha_{\rm AAK}$ as a function of $D$.
The dashed line represents $D^{-1/3}$.}
\label{1_tau_phi_coeff_vs_D}
\end{center}
\end{figure}

To make a quantitative analysis, we plot in
Fig.~\ref{1_tau_phi_coeff_vs_D} the experimental parameter $a_{\rm
exp}$ of Eq.~(\ref{AAK_fitting_1D}), normalized by the theoretical
prefactor $\alpha_{\rm AAK}$ of Eq.~(\ref{AAK_alpha}), as a function
of $D$.~\cite{note_definition_D} In the diffusive regime, the parameter $a_{\rm
exp}/\alpha_{\rm AAK}$ follows a power law as a function of $D$ with
$a_{\rm exp}/\alpha_{\rm AAK}\propto D{^{-1/3}}$, which is
consistent with Eq.~(\ref{AAK_alpha}). Moreover, the prefactor
$a_{\rm exp}$ obtained in this work agrees with Eq.~(\ref{AAK_1D})
in absolute value within $15\%$. In the semi-ballistic regime, on
the other hand, we obtain a very different behavior of $a_{\rm
exp}/\alpha_{\rm AAK}$ as a function of $D$. While in the diffusive
regime the parameter $a_{\rm exp}/\alpha_{\rm AAK}$ is in accordance
with the diffusive theory, in the semi-ballistic regime the
decoherence time seems to be \textit{independent} of the disorder.
On the other hand, we observe the same width dependence of prefactor
$a_{\rm exp} \sim w^{-2/3}$ as in  the diffusive regime. From these
experimental facts, it is obvious that the temperature and width
dependence of the phase coherence time $\tau_{\phi}$ in the
semi-ballistic regime are well captured within the AAK theory,
whereas the disorder dependence of $\tau_{\phi}$ has to be
reconsidered in the semi-ballistic regime.

\begin{figure}
\begin{center}
\includegraphics[width=6cm]{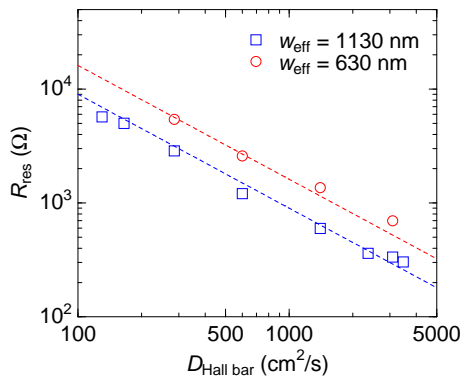}
\caption{(Color online) Residual resistance of the wires as a function of $D$ obtained from the Hall bar. The broken lines (blue and red) show $R_{\rm res} \propto 1/Dw_{\rm eff}$ for $w_{\rm eff}=$ 1130 and 630 nm, respectively.
}
\label{R_residual_vs_D}
\end{center}
\end{figure}

One could argue that the \textit{disorder-independent} decoherence
time in the semi-ballistic regime might be simply due to saturation
of the diffusion coefficient $D$. If the boundary scattering in
quasi-1D wires were diffusive, the diffusion coefficient should
saturate at $D=1/2(v_{F}w_{\rm eff})$,~\cite{Beenakker_prb_88} which
could lead to a $D$-\textit{independent} $\tau_{\phi}$. This
possibility, however, can be ruled out by plotting the resistance of
the wires as a function of $D$. Figure~\ref{R_residual_vs_D} shows
the residual resistance of the quasi-1D wires $R_{\rm res}$ [see
Eq.~(\ref{residual_resistance}) in Sec. VI] as a function of $D$
obtained from the Hall bar.~\cite{note_residual_resistance}
The residual resistance $R_{\rm res}$
nicely follows a $1/Dw_{\rm eff}$ law over the whole $D$ range [see
Table~\ref{table1}]. This dependency can be realized only when the
boundary scattering in the semi-ballistic regime is specular.
Moreover, as mentioned in Sec. IV B, our wires have been made by
shallow etching which results in highly specular boundary
reflection.~\cite{shallow_etching} The $D$ dependence of the
residual resistance also confirms our assumption that
$L_{\phi}=\sqrt{D \tau_{\phi}}$ is valid even in the semi-ballistic
regime.

To our knowledge, there is no theoretical prediction about the
disorder dependence of the decoherence for quasi-1D wires in the
clean limit (very few impurities). There are, however, a few
theoretical works to give us some hints. It should be noted that
these calculations have been performed for 2D systems. Wittmann and
Schmid calculated the 2D WL correction for arbitrary number of
elastic scattering time $\tau_{e}$.~\cite{wittmann_jltp_87} They
found that the WL correction in the clean limit can be reduced
compared to the diffusive case, leading to an under-estimation of
$\tau_{\phi}$. Narozhny and co-workers calculated the temperature
dependence of $\tau_{\phi}$ in a 2D system at arbitrary relation
between $k_{B}T$ and $\hbar/\tau_{e}$.~\cite{aleiner_prb_02} They
showed that the phase coherence time $\tau_{\phi}$ has the same
temperature dependence both in the diffusive and ballistic regimes,
but the prefactor in the ballistic regime is smaller than in the
diffusive one. These theoretical calculations are qualitatively
consistent with our experimental result on the quasi-1D wires; as is
shown in Fig.~\ref{1_tau_phi_coeff_vs_D}, the dephasing time
$\tau_{\phi}$ in the semi-ballistic regime is independent of $D$
while $\tau_{\phi}$ in the diffusive regime is quantitatively
consistent with the AAK theory, i.e. $\tau_{\phi}\propto D^{1/3}$.
However, it is not possible to make a quantitative analysis of the
diffusion coefficient dependence of $\tau_{\phi}$ on the basis of
these calculations. It is desirable that theoretical calculations of
$\tau_{\phi}$ in the semi-ballistic regime are performed for the
quasi-1D wires.

\subsection{Comparison with theory on zero temperature decoherence}

As pointed out in the introduction, decoherence in metallic systems
at zero temperature has been a controversial issue over the last
decade.~\cite{AAK_82,mohanty_prl_97,GZ_prl_98,GZ_prb_06,imry_epl_99,zawa_prl_99,birge+pierre_prl_02,schopfer_prl_03,pierre_prb_03,bauerle_prl_05,birge_prl_06,mallet_prl_06,sami_physicaE_07,capron_prb_08,glazman_03,zarand_prl_04,Borda_07,rosch_prl_06,Micklitz_07,Costi_prl_08,gershenson_prl_06,Lin_2001,Noguchi_JAP_96,niimi}
By studying only the temperature
dependence of the phase coherence time it is very difficult to
discriminate experimentally whether a saturating decoherence time is
observed or not. Firstly, several precautions have to be taken such
that an experimentally observed saturation is not caused by either
external radio frequency propagating along the measuring lines or by
the determination of the actual electron temperature of the sample
which is not always straightforward. Secondly, even if all these
requirements are fulfilled, a small inclusion of magnetic impurities
will always lead to a saturating decoherence time at very low but
finite temperature.~\cite{bauerle_prl_05,mallet_prl_06,sami_physicaE_07}
In addition, to avoid magnetic impurities in metallic systems is extremely
difficult as metallic sources cannot be purchased with a guaranteed
impurity level below the ppm level. It is hence clear that simply
studying the temperature dependence is not sufficient to give a
definite answer to the saturation problem.
A different approach to this problem can be done by studying the
diffusion coefficient dependence of the decoherence time.
Compared to the AAK theory, the GZ theory predicts a much stronger
diffusion coefficient dependence of $\tau_{\phi}$
at very low temperatures~\cite{GZ_physica_E_07} as detailed below.
This can be tested with the present experiment.

\begin{figure}
\begin{center}
\includegraphics[width=6cm]{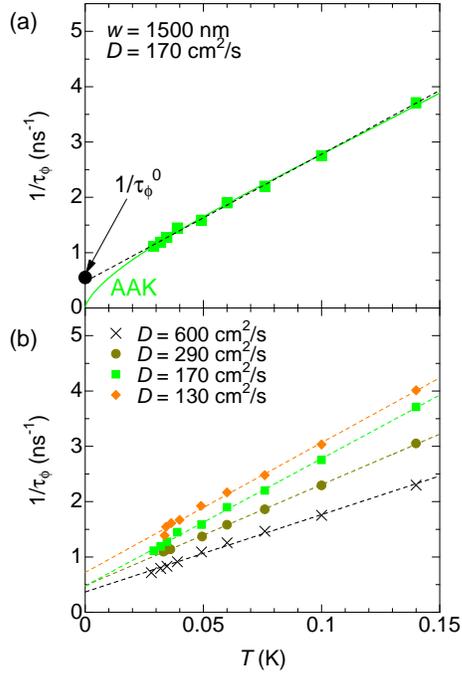}
\caption{(Color online) (a) An example to extract $1/\tau_{\phi}^{0}$ proposed by
GZ.~\cite{GZ_jltp_02} The solid line shows the best fit of the AAK
formula Eq.~(\ref{AAK_1D_1}). The broken line is a linear fit of the
dephasing rate $1/\tau_{\phi}(T)$. The intercept represents
$1/\tau_{\phi}^{0}$. (b) Dephasing rate $1/\tau_{\phi}$ at different
$D$ as a function of $T$ on a linear scale ($w=1500$ nm). The broken
lines have the same meaning as in (a).}
\label{dephasing_time_1500nm}
\end{center}
\end{figure}

\begin{figure}
\begin{center}
\includegraphics[width=6.5cm]{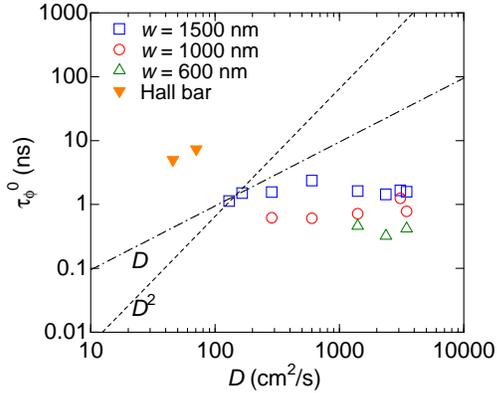}
\caption{(Color online) $\tau_{\phi}^{0}$ for all the investigated samples. The
dotted and dashed-dotted lines show $D^{2}$ and $D$ laws,
respectively.} \label{GZ_tau_phi_0_all}
\end{center}
\end{figure}

According to the GZ theory, $\tau_{\phi}(T)$ intrinsically saturates
at zero temperature in the ground state of a disordered conductor at
a finite value $\tau_{\phi}^{0}$ due to the fluctuations of the
electromagnetic field generated by an electron and which is
experienced by the other electrons.~\cite{GZ_prl_98,GZ_prb_06} The
finite value depends strongly on the intrinsic disorder. In
particular the GZ theory predicts that $\tau_{\phi}^{0} \propto
D^{2}$ for 2D and $\tau_{\phi}^{0} \propto D$ for
1D.~\cite{GZ_prb_99,GZ_jltp_02} Note that the dimensionality here is
determined in terms of $l_{e}$; the former case should be applied in
the diffusive regime where $L,w>l_{e}$, while the latter case should
be applied in the semi-ballistic regime where
$L>l_{e}>w$.~\cite{GZ_metallic_wire} The AAK theory, on the other
hand, predicts a very slow $D$ dependence of the dephasing time,
i.e. $\tau_{\phi}\propto D^{1/3}$, as shown in Eq.~(\ref{AAK_1D}).

The fact that we do not see any apparent saturation in the
temperature dependence of $\tau_{\phi}$ or $L_{\phi}$ for all
samples investigated [see Fig.~\ref{L_phi_all}] seems already in
contradiction with the GZ theory. Nevertheless, we will adopt the
method proposed in Ref.~\onlinecite{GZ_jltp_02} to extract the saturation
time $\tau_{\phi}^{0}$. This can be done by plotting the inverse of
the dephasing time (dephasing rate) as a function of temperature on
a linear scale. By extrapolating a linear fit to the low temperature
data down to zero temperature (in our case we take all the data
below 150 mK for the fitting), one obtains $\tau_{\phi}^{0}$ as
shown in Fig.~\ref{dephasing_time_1500nm}. For comparison we also
plot the theoretical expectation within the AAK theory. We then
determine $\tau_{\phi}^{0}$ in the same way for all diffusion
coefficient investigated. This is shown in
Fig.~\ref{GZ_tau_phi_0_all} for three wires with different width as
well as for the Hall bars. For our data we obtain a very weak
variation of $\tau_{\phi}^{0}$ as a function of diffusion
coefficient. It is clear that the diffusion coefficient dependence
of $\tau_{\phi}^{0}$ is much weaker than the one expected within the
GZ theory (dotted and dashed-dotted lines). One could of course
argue that our measurements do not extend to low enough temperature
and that the saturation of $\tau_{\phi}$ will only occur at lower
temperature. This contrasts however with the fact that for metals
with similar diffusion coefficients very frequently a saturation of
$\tau_{\phi}$ is observed at much higher temperatures. These facts
therefore suggest that the frequently observed low temperature
saturation of $\tau_{\phi}$ is \textit{not} intrinsic.

\section{Temperature dependence of the resistance}

As mentioned above, an important issue in this paper is decoherence
at zero temperature. For decoherence measurements at very low
temperatures, it is important to know the actual electron
temperature of the sample which can be quite different than that of
the thermal bath. In order to probe the electron temperature of the
2DEG \textit{in situ}, we have used the temperature dependence of
the Altshuler-Aronov correction term as detailed in the following
subsection.

\subsection{Altshuler-Aronov correction}

In the diffusive regime, the electrical resistance of a quantum wire
(or Hall bar) consists of different contributions:
\begin{eqnarray}
R(B,T) &=& R_{\rm res} + R_{\rm e-ph} (T) \nonumber\\
&+& \Delta R_{\rm WL} (B, T) + \Delta R_{\rm AA} (T) +  \cdots.
\label{residual_resistance}
\end{eqnarray}
The first term $R_{\rm res}$ corresponds to
the residual resistance and the second term
comes from the e-ph interactions. At high temperatures $R_{\rm
e-ph}$ simply follows a $T$-linear dependence and vanishes as
temperature goes to zero. The third term is the WL quantum
correction term which has already been described in Sec. III A. The
last term is the so-called Altshuler-Aronov (AA)
correction.~\cite{AA_correction} At low temperatures, the e-e
interactions are responsible for a small depletion of the density of
states at the Fermi energy which leads to a correction to the
resistivity. Basically, the WL and AA corrections are of the same
order, but the latter can be distinguished from the former by
applying a small magnetic field which suppresses the WL correction.
The AA correction in the quasi-1D case is given as below:
\begin{eqnarray}
\Delta R_{\rm AA} (T) &\equiv& R(T)-R_{\rm res} \nonumber \\
&=& 0.782\lambda_{\sigma} R_{\rm res}^{2}N\frac{e^{2}}{h}\frac{L_{\rm T}}{L} \nonumber \\
&=& R_{\rm res}^{2}\frac{A_{\rm theo}}{\sqrt{T}}. \label{AA_correction_1D}
\end{eqnarray}
The parameter $\lambda_{\sigma}$ is a constant which represents the
strength of the screening of the interactions. In the quasi-1D case,
one has $\lambda_{\sigma}=4-3F/2$ where $F$ is the screening factor
varying from 0 for an unscreened interaction to 1 for a perfectly
screened interaction. In a similar manner, one can obtain the 2D AA
correction in the limit $T<\hbar/(k_{B}\tau_{e}$):
\begin{eqnarray}
\Delta R_{\rm AA} (T) = \lambda_{\sigma} R_{\rm res}^{2}\frac{e^{2}}{2\pi h}\frac{w}{L} \left\{ \gamma - \ln \left(\frac{2\pi k_{B}T \tau_{e}}{\hbar} \right) \right\} \label{AA_correction_2D}
\end{eqnarray}
where $\gamma \simeq 0.577$ is the Euler constant
and $\lambda_{\sigma}=2-3F/2$.


\subsection{Experimental results in the diffusive regime}

\begin{figure}
\begin{center}
\includegraphics[width=6.5cm]{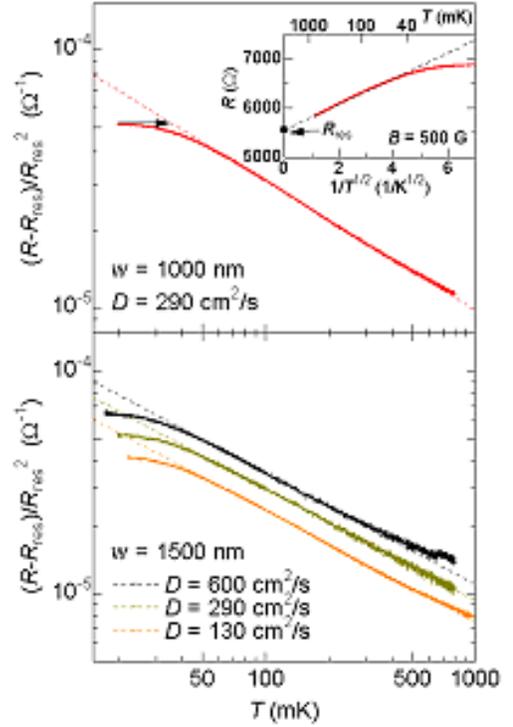}
\caption{(Color online) Resistance variations of 1000 (top) and 1500 nm wide wires (bottom) as a function of $T$ at $B=500$ and 300 G, respectively. The broken lines are the best fits of Eq.~(\ref{AA_correction_1D}). The \textit{real electron temperature} below 40 mK is corrected by the $1/\sqrt{T}$ law (see the arrow). In the inset of the top figure, the resistance of 1000 nm wide wires is plotted as a function of $1/\sqrt{T}$ to extract $R_{\rm res}$.}
\label{R_vs_T_3ppm}
\end{center}
\end{figure}

At fields high enough to suppress the WL correction ($B
= 150 \sim 500$ G), the resistance of a quasi-1D metallic wire
follows a $1/\sqrt{T}$ law due to electron-electron interactions and
can be used as a ``thermometer" to probe the \textit{effective
electron temperature}~\cite{bauerle_prl_05}. For this purpose, we
plot the resistance of our 1000 nm wide wire as a function of
$1/\sqrt{T}$ in the inset of the top figure of
Fig.~\ref{R_vs_T_3ppm}. It follows nicely the $1/\sqrt{T}$
dependence down to 40 mK~\cite{note_L_T}. Below this temperature it
starts to deviate from the $1/\sqrt{T}$ law. This is also observed
for wires with different widths and different diffusion
coefficients. To show the deviation more clearly, $(R-R_{\rm
res})/R_{\rm res}^{2}$ is plotted as a function of temperature in
Fig.~\ref{R_vs_T_3ppm}, where $R_{\rm res}$ is obtained by
extrapolating the $R$ vs $1/\sqrt{T}$ curve down to zero [see inset
of Fig.~\ref{R_vs_T_3ppm}]. Assuming that the $1/\sqrt{T}$
dependence of the resistance holds down to the lowest temperature,
we obtain an effective electron temperature of 25 mK at the base
temperature of our cryostat. This fact is also confirmed by the
temperature dependence of the phase coherence length [see
Fig.~\ref{L_phi_D_290}]. Therefore, all our data have been
temperature corrected below 40 mK.

\begin{figure}
\begin{center}
\includegraphics[width=6cm]{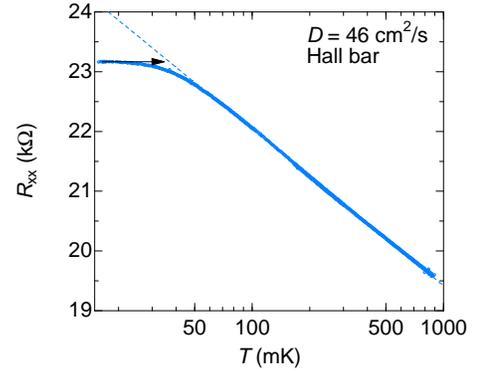}
\caption{(Color online) Resistance variation of the Hall bar at $D=46$ cm$^{2}$/s as a function of $T$. The broken line shows a $\ln(T)$ law.}
\label{R_vs_T_15ppm}
\end{center}
\end{figure}

In Fig.~\ref{R_vs_T_15ppm}, the resistance variation of the 2D Hall
bar for $D = 46$ cm$^{2}/$s is plotted as a function of $T$ on a
semi-log scale. As expected from Eq.~(\ref{AA_correction_2D}), the
AA correction term follows a $\ln (T)$ law down to 40 mK. Like in
the case of quasi-1D wires, below this temperature the resistance
deviates from the theoretical expression. In a similar manner we
correct the actual temperature below 40 mK.

\subsection{Experimental results in the semi-ballistic regime}

\begin{figure}
\begin{center}
\includegraphics[width=7cm]{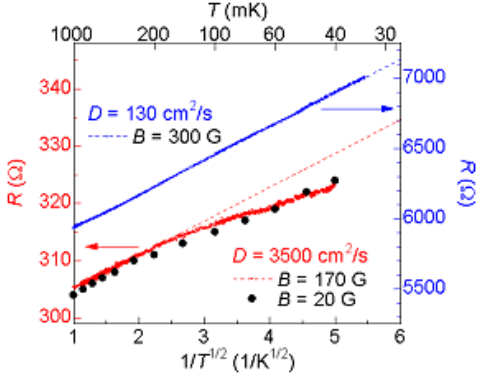}
\caption{(Color online) Resistance of 1500 nm wide wires as a
function of $1/\sqrt{T}$ at $D=3500$ (left axis) and 130 cm$^{2}$/s
(right axis).  The resistance at $D=130$ cm$^{2}$/s is measured at
$B=300$ G ($B/B^{*} = 0.07$), while the resistance at $D=3500$
cm$^{2}$/s is measured at $B=20$ ($B/B^{*} = 0.12$; black dots) and
170 G ($B/B^{*} = 1$; red curve). The broken lines are linear fits
to extract $R_{\rm res}$. The temperature below 40 mK has already
been corrected for both diffusion coefficients.}
\label{R_vs_T_ballistic_diffusive}
\end{center}
\end{figure}

In the semi-ballistic regime where $l_{e}>w_{\rm {eff}}$, we find an
unexpected temperature dependence of the resistance. In
Fig.~\ref{R_vs_T_ballistic_diffusive}, a resistance vs $1/\sqrt{T}$
curve in this regime ($D=3500$ cm$^{2}$/s) is compared to that in
the diffusive regime (130 cm$^{2}$/s). As discussed above, in the
diffusive regime and at fields high enough to suppress WL the
resistance follows nicely a $1/\sqrt{T}$ law in the entire
temperature range. In the semi-ballistic regime, on the other hand,
we observe a deviation from the $1/\sqrt{T}$ law below 150 mK which
is somewhat unexpected.

In this regime one has to be careful about the applied magnetic field to
suppress WL such that it does not affect the trajectories of the electrons, 
in other words, does not lead the SdH oscillations.
According to Ref.~\onlinecite{gornyi_prb_04}, the AA correction to
resistance is independent of $B$ when the condition $B/B^{*} \ll 1$
is satisfied. We have therefore measured the e-e interaction
correction for different magnetic fields as shown in
Fig.~\ref{R_vs_T_ballistic_diffusive}. For fields lower than 170 G
($B/B^{*} = 1$) we do not observe a significant change in the
temperature dependence and we can rule out the possibility 
that the observed temperature dependence 
is due to the applied magnetic field. It is
also unlikely that the observed temperature dependence is due to a
decoupling of the electrons from the thermal bath since 
the phase coherence length nicely follows the AAK theory down to
the lowest temperatures as shown in Fig.~\ref{L_phi_all}. 
We also exclude the possibility that this temperature
dependence results from a dimensional crossover when the thermal
length $L_{T} = \sqrt{\hbar D /k_{B} T}$ becomes comparable to the
width of the wire $w_{\rm eff}$.~\cite{note_L_T}

When entering the semi-ballistic regime $(l_e > w_{\rm eff})$, as the
scattering at the boundaries in our wires is mostly specular, the
temperature dependence of the e-e interactions may be
influenced~\cite{zaikin_prb_04,mora_prb_07} and modified by an
additional logarithmic term at intermediate temperatures 
($k_{B}T\tau_{e}/\hbar \approx 1$).

\begin{figure}
\begin{center}
\includegraphics[width=6.5cm]{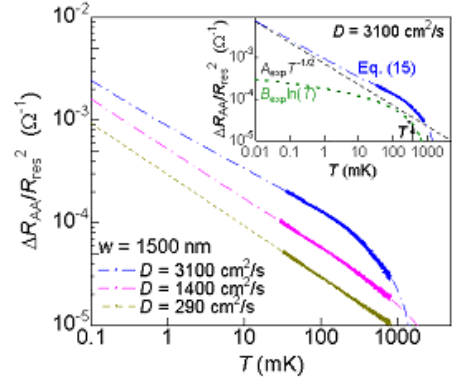}
\caption{(Color online) Resistance variations of 1500 nm wide wires
as a function of $T$ for three different diffusion constants. The
dashed-dotted lines are the best fits to
Eq.~(\ref{AA_correction_SB}). Inset: Resistance variation of the
1500 nm wide wires at $D=$3100 cm$^{2}/$s. The broken line (black)
shows the $1/\sqrt{T}$ law whereas the dotted line (green) shows
the $\ln(T)$ dependence. The dashed dotted line (blue) is again the
best fit to Eq.~(\ref{AA_correction_SB}). $T^*$ indicates the
temperature where $T^{*} = \hbar/(\tau_{e}k_{B})$.} \label{R_vs_T_1500nm_different_D}
\end{center}
\end{figure}

\begin{figure}
\begin{center}
\includegraphics[width=7cm]{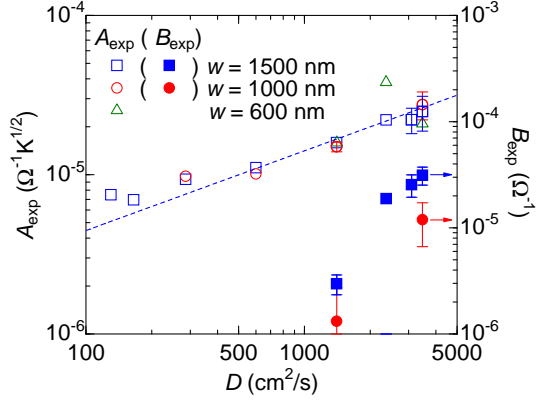}
\caption{(Color online) The experimental prefactors $A_{\rm exp}$ (left axis) and $B_{\rm exp}$ (right axis) of the AA correction as a function of $D$. 
The broken line shows a $D^{1/2}$ law.} \label{prefactor_e-e_vs_D}
\end{center}
\end{figure}

In the following, we will try to fit the observed temperature
dependence of the e-e interaction correction by a combination of a
$1/\sqrt{T}$ and a logarithmic term:
\begin{eqnarray}
\frac{\Delta R_{\rm AA} (T)}{R_{\rm res}^{2}}&=& \frac{A_{\rm
exp}}{\sqrt{T}} +B_{\rm exp} \ln (T).
\label{AA_correction_SB}
\end{eqnarray}
This is shown in Fig.~\ref{R_vs_T_1500nm_different_D}. Indeed,
fitting with Eq.~(\ref{AA_correction_SB}) reproduces fairly well the
observed temperature dependence in the semi-ballistic regime 
[see dashed-dotted lines in Fig.~\ref{R_vs_T_1500nm_different_D}]. 
Deep in the semi ballistic regime we see a relatively strong deviation
from the $1/\sqrt{T}$ dependence. By decreasing the
diffusion coefficient, the temperature dependence becomes more and
more 1D like and turns completely into the 1D regime when entering
the diffusive regime ($l_{e} < w_{\rm eff}$). From fitting the data with
Eq.~(\ref{AA_correction_SB}) we can extract the values of the
prefactors of the 1D $(A_{\rm exp})$ as well as logarithmic behavior
$(B_{\rm exp})$ as shown in Fig.~\ref{prefactor_e-e_vs_D}. We observe
that the prefactor of the 1D contribution is proportional to
$D^{1/2}$ as expected from Eq.~(\ref{AA_correction_1D}). In
addition, $A_{\rm exp}$ shows no wire width dependence which is
consistent with Eq.~(\ref{AA_correction_1D}). In the diffusive
regime ($D<1000$ cm$^2$/s), the logarithmic contribution is
negligible. However, when entering the semi-ballistic regime, the
prefactor of the logarithmic contribution becomes comparable to the
1D term and dominates the 1D term for our cleanest samples. In the
overall temperature dependence, the additional logarithmic
contribution shifts the crossover temperature where the 1D AA
behavior dominates to much lower temperatures. This is in line with
the crossover calculated in Ref.~\onlinecite{mora_prb_07} 
where the crossover temperature $T^{*}$ is renormalized due to the
electron-electron interactions.

\section{Strongly localized regime}

So far, we have discussed decoherence in the weakly localized regime
for quasi-1D wires and 2D Hall bars. In that regime, one has to meet
conditions such that the $k_{F}l_{e}$ value is much larger than 1
and also the localization length $\xi_{\rm loc}$ is much larger than
$L_{\phi}$. By increasing the disorder, however, one can reach a
regime where $k_{F}l_{e}$ is of the order of 1 and which is usually
referred to as the \textit{strongly localized} regime. In this last
section we will present measurements of the resistance as well as
the phase coherence length in quasi-1D wires and 2D Hall bars in this
regime.

\subsection{2D Hall bars}

For the 2D case a fair amount of
experimental~\cite{faran_prb_88,hsu_prl_95,tremblay_jp_90,jiang_prb_92,keuls_prb_97,khondaker_prb_99,minkov_prb_02,minkov_prb_04,minkov_prb_07,peled_prl_03,li_prl_09}
as well as theoretical
works~\cite{efros_jp_75,vollhardt_prl_80,mott_jp_81,davies_jp_82,gogolin_ssc_93,zhao_prb_91}
can be found in the literature. It is commonly believed that the
conduction process in the strongly localized regime is attributable
to 2D variable
range hopping, 
and several experiments support this
assumption.~\cite{tremblay_jp_90,jiang_prb_92,keuls_prb_97,khondaker_prb_99}
On the contrary, the question on how decoherence is affected when
going from the \textit{weakly localized} to the \textit{strongly
localized} regime is still open.

This problem has been studied mainly in semiconductor
heterojunctions with
2DEGs.~\cite{tremblay_jp_90,jiang_prb_92,keuls_prb_97,khondaker_prb_99,minkov_prb_02,minkov_prb_04,minkov_prb_07,peled_prl_03,li_prl_09}
In such 2D systems, an estimation of the localization length
$\xi_{\rm loc}^{\rm 2D}$ is given by:~\cite{minkov_prb_04,minkov_prb_07}
\begin{eqnarray}
\xi_{\rm loc}^{\rm 2D} &=& l_{e}\exp\left(\frac{\pi}{2} k_{F}l_{e}\right) \nonumber\\
&=&\frac{2\sqrt{2\pi}m^{*}}{\sqrt{n}h}D\exp\left(\frac{2\pi^{2}m^{*}}{h}D\right).
\label{localization_length_2D}
\end{eqnarray}
When $\xi_{\rm loc}^{\rm 2D}$ becomes comparable or smaller than the
phase coherence length $L_{\phi}$, one enters the strongly localized
regime.

\begin{figure}
\begin{center}
\includegraphics[width=6cm]{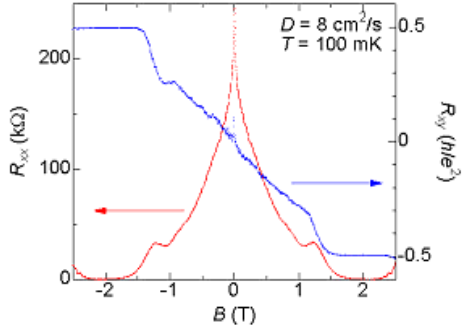}
\caption{(Color online) $R_{xx}$ and $R_{xy}$ at $D=8$ cm$^{2}/$s and $T=100$ mK.
The Hall resistance $R_{xy}$ is normalized by $h/e^{2}$.}
\label{Ga30ppm_Hall_bar}
\end{center}
\end{figure}

\begin{figure}
\begin{center}
\includegraphics[width=7.0cm]{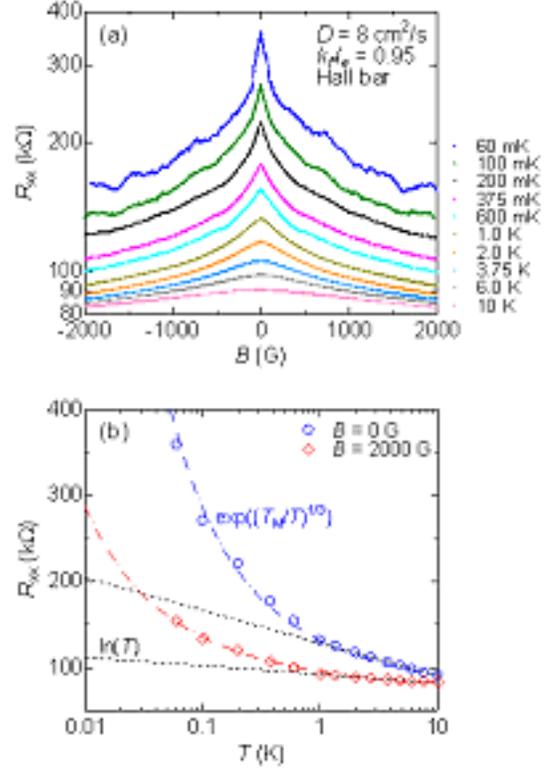}
\caption{(Color online) (a) Magnetoresistance curves of the Hall bar at $D=8$
cm$^{2}/$s at several different temperatures. (b) $R_{xx}$ at $B=0$
and 2000 G as a function of temperature. The broken and
dashed-dotted lines represent $\ln(T)$ and 2D variable range hopping
laws, respectively.} \label{Ga30ppm_WL}
\end{center}
\end{figure}

In Fig.~\ref{Ga30ppm_Hall_bar} we show $R_{xx}$ and $R_{xy}$ at
$k_{F}l_{e}=0.95$ (or $D=8$ cm$^{2}$/s) at $T=100$ mK. At $B \sim 2$
T, we can still observe the $\nu=2$ quantum plateau where
$R_{xx}=0$. At low fields, $R_{xx}$ shows a large negative
magnetoresistance which is more than 10 times larger than $h/e^{2}$
for $B=0$. In order to see how $R_{xx}$ evolves with temperature in
the low field region, we plot the magnetoresistance for different
temperatures on a semi-log plot in Fig.~\ref{Ga30ppm_WL}(a). With
decreasing temperature, the peak height exponentially grows but the
shape of the magnetoresistance seems to be similar to that in the
weakly localized regime down to $T \sim 100$ mK [see
Fig.~\ref{WL_curves_D_46}]. Below this temperature, $R_{xx}$ near
zero field is extremely enhanced. Such a \textit{large} negative
magnetoresitance is probably a precursor of the $\exp(-\sqrt{B})$
law expected in the coherence interference
model.~\cite{zhao_prb_91} Let us now discuss in more detail the
temperature dependence of the resistance at zero field and at a
field of 2000 G where the WL correction is basically suppressed. As
seen in Fig.~\ref{Ga30ppm_WL}(b), above 1 K $R_{xx}$ follows a
$\ln(T)$ dependence as expected in the weakly localized regime [see
Fig.~\ref{R_vs_T_15ppm}]. Below 1 K, $R_{xx}$ deviates from the
$\ln(T)$ law and can be fitted to a 2D variable range hopping law
$R(T) \propto \exp(T_{M}/T)^{1/3}$; $T_{M}=300$ and 28 mK for 0 and
2000 G, respectively. Such a behavior has already been seen in other
experiments in the strongly localized
regime.~\cite{tremblay_jp_90,jiang_prb_92,keuls_prb_97,khondaker_prb_99}

\begin{figure}
\begin{center}
\includegraphics[width=7.0cm]{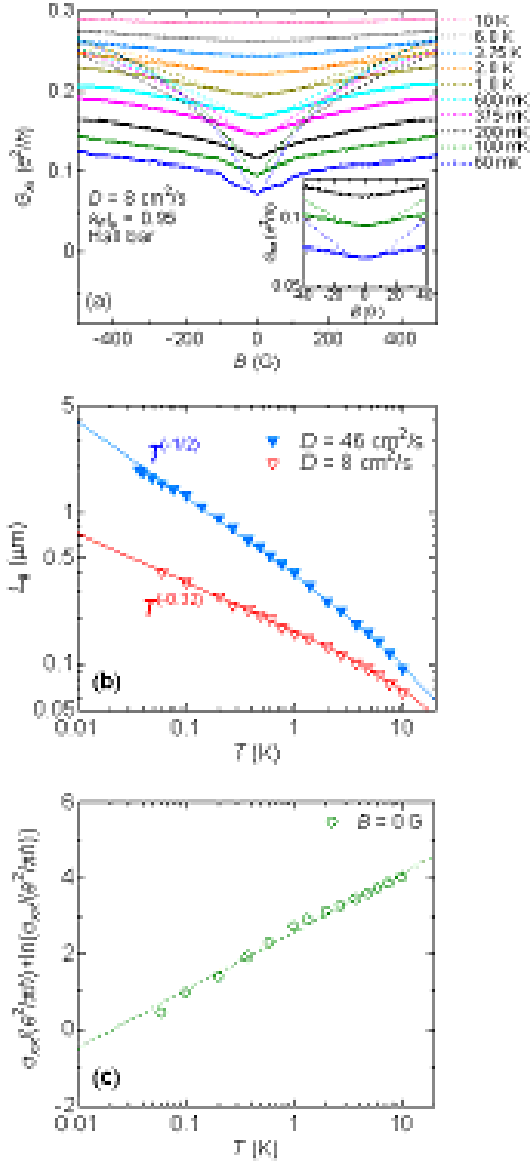}
\caption{(a) Magnetoconductance curves of the Hall bar at $D=8$
cm$^{2}/$s at several different temperatures. The conductance is
divided by $e^{2}/h$. The broken lines show the best fits of
Eq.~(\ref{HLN_2D}). The inset shows a closeup view of the low field
part of the magnetoconductance at low temperatures. (b) Temperature
dependence of $L_{\phi}$ in the strongly localized regime (open
symbols). For comparison, $L_{\phi}$ in the weakly localized regime
(closed symbols) is also plotted. (c) Temperature dependence of
$\sigma_{xx}$ in the representation corresponding to the
self-consistent theory Eq.~(\ref{self-consistent_theory}).}
\label{Ga30ppm_L_phi}
\end{center}
\end{figure}

As pointed out above, the shape of the magnetoresistance is similar
to that in the weakly localized regime. Although the WL theory
Eq.~(\ref{HLN_2D}) is in principle only applicable in the weakly
localized regime, we nevertheless fit the magnetoconductance curves
with Eq.~(\ref{HLN_2D}) as shown in Fig.~\ref{Ga30ppm_L_phi}(a) at
temperatures higher than 60 mK. A similar approach has already been
done by Minkov \textit{et al.}~\cite{minkov_prb_02,minkov_prb_07} Let us recall
that Eq.~(\ref{HLN_2D}) is limited to a small field range within
$B_{c}=\hbar/4eL_{\phi}^{2}$. At high temperatures, the fitting
works very well in a relatively wide field range. Going to lower
temperatures, the fitting region is getting smaller which indicates
that $L_{\phi}$ increases. The obtained $L_{\phi}$ from the WL
theory is plotted as a function of $T$ in
Fig.~\ref{Ga30ppm_L_phi}(b). The phase coherence length $L_{\phi}$
of the Hall bar in the \textit{strongly} localized regime follows a
power law $T^{p}$  at low temperatures as indicated by the solid
line, just like in the weakly localized regime, but with a smaller
exponent $p=-0.32$. Such a temperature dependence is very similar to
what has been observed in Ref.~\onlinecite{minkov_prb_04} for similar
values of $k_{F} l_{e}$. In that work,~\cite{minkov_prb_04} the
exponent varied from $p=-0.5$ to $-0.3$ when reducing $k_{F} l_{e} <
5$ down to $k_{F} l_{e} \sim 1$, similar to our observations.

Within the theoretical approach of the phase coherence in the
Anderson localization regime proposed by Vollhardt and
W\"{o}lfle,~\cite{vollhardt_prl_80,gogolin_ssc_93} the conductivity can be
calculated for arbitrarily weak disordered systems.
Their self-consistent theory leads to the following
equation for the conductivity $\sigma_{xx}
(T)$:~\cite{minkov_prb_02}
\begin{eqnarray}
\left[\frac{\sigma_{xx}(T)}{(e^{2}/\pi h)}+\ln \left( \frac{\sigma_{xx}(T)}{e^{2}/\pi h}\right) \right] &=& \pi k_{F} l_{e} +\ln (\pi k_{F}l_{e}) \nonumber\\
&-&2\ln \left(\frac{L_{\phi}(T)}{l_{e}} \right),
\label{self-consistent_theory}
\end{eqnarray}
where we assume that $L_{\phi}=\sqrt{D\tau_{\phi}}$ and $L_{\phi}
\gg l_{e}$. Strictly speaking, Eq.~(\ref{self-consistent_theory}) is
valid only when $k_{F}l_{e} \gg 1$. Nevertheless, inspired by
Ref.~\onlinecite{minkov_prb_02}, we plot the left side of
Eq.~(\ref{self-consistent_theory}) for $B=0$ G as a function of $T$
in Fig.~\ref{Ga30ppm_L_phi}(c). It exhibits a $\ln(T)$ dependence
over the whole temperature range.~\cite{note_self_consistent_theory}
Such a $\ln(T)$ law is expected if one assumes a power law for the
temperature dependence of the phase coherence length. From the slope
of the left side of Eq.~(\ref{self-consistent_theory}) vs $T$ curve,
we can determine the exponent of $L_{\phi} (T)$ ($L_{\phi}(T)
\propto T^{p}$). Interestingly, we again obtain $p \simeq -0.32$
which is identical to the one obtained when fitting the temperature
dependence of the magnetoconductance with the WL theory [see
Fig.~\ref{Ga30ppm_L_phi}(b)]. This hints to the conclusion that when
going from the weakly localized to the strongly localized regime the
temperature dependence of $L_{\phi}$ is still diverging with
decreasing temperature with a power law, but with a smaller exponent
compared to the weakly localized regime.

\begin{figure}
\begin{center}
\includegraphics[width=5.5cm]{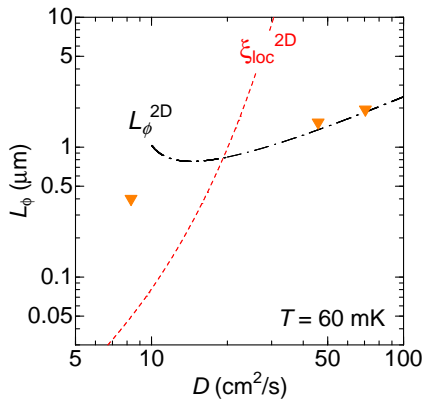}
\caption{(Color online) Phase coherence length $L_{\phi}$ of Hall bars
at $T=60$ mK as a function of $D$. The broken and dashed-dotted lines represent
the 2D localization length Eq.~(\ref{localization_length_2D}) and
theoretically expected phase coherence length $L_{\phi}^{2D}$
based on Eq.~(\ref{AAK_2D_1}), respectively.}
\label{tau_N_2D}
\end{center}
\end{figure}

Before closing this subsection, let us mention the diffusion coefficient
dependence of the phase coherence length in 2D systems.
In Fig.~\ref{tau_N_2D}, we plot $L_{\phi}$ obtained at $T=60$ mK
in 2D Hall bars as a function of $D$.
In the weakly localized regime, 
$L_{\phi}$ nicely follows the formula based on Eq.~(\ref{AAK_2D_1})
as shown in the dashed-dotted line in Fig.~\ref{tau_N_2D}. With
decreasing $D$, this formula diverges because of the logarithmic
term in Eq.~(\ref{AAK_2D_1}),~\cite{note_L_N_2D} and the 2D
localization length $\xi_{\rm loc}^{\rm 2D}$ becomes smaller than
the phase coherence length. In the strongly localized regime at zero
temperature electrons should be localized within a length scale of
$\xi_{\rm loc}^{\rm 2D}$. At finite temperatures, on the other hand,
they can hop from one island with a size of $\xi_{\rm loc}^{\rm 2D}$
to another, and this hopping process gives rise to the exponential
increase of the resistance as shown in Fig.~\ref{Ga30ppm_WL}(b).
During this process the phase coherence of the electrons should be
maintained within a length scale of $L_{\phi}$. Thus, in the
strongly localized regime, the phase coherence length $L_{\phi}$ can
be larger than the localization length $\xi_{\rm loc}^{2D}$.

\subsection{1D wires}

\begin{figure*}
\begin{center}
\includegraphics[width=17cm]{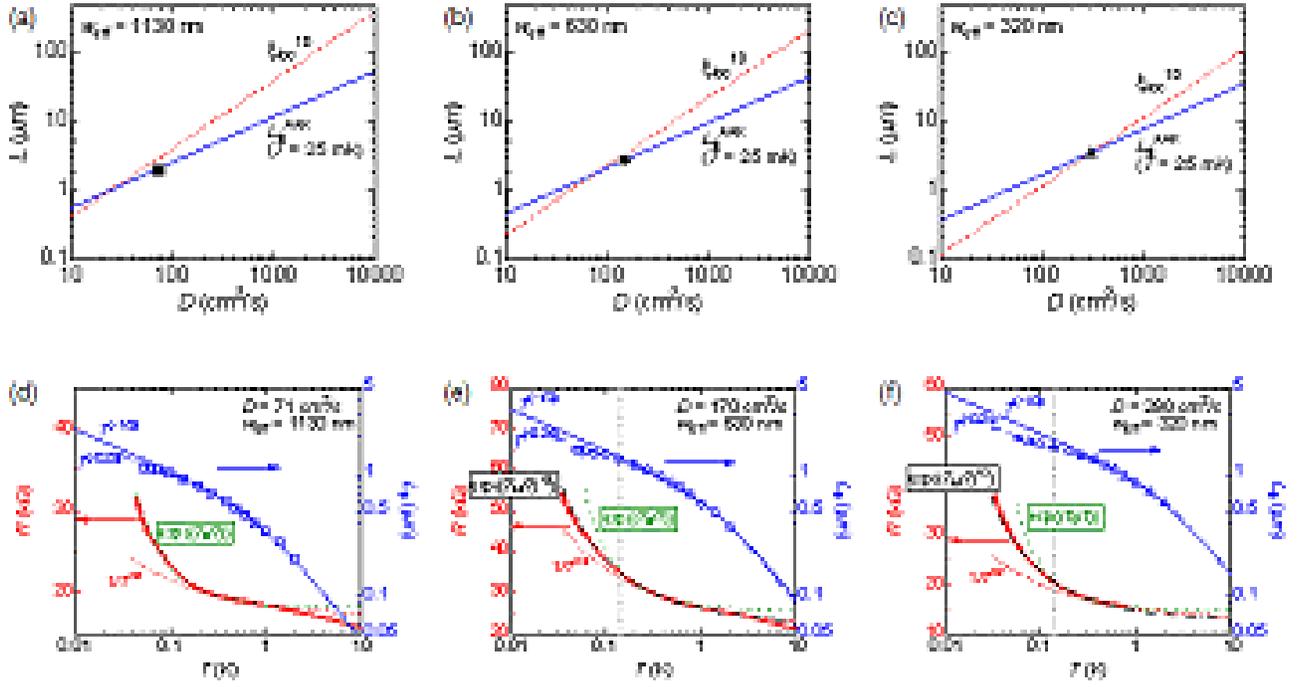}
\caption{(Color online) (a)-(c) Diffusion coefficient dependence of the theoretical
1D localization length Eq.~(\ref{localization_length_1D}) and phase
coherence length expected in the AAK theory at $T=25$ mK; (a)
$w_{\rm eff}=1130$, (b) 630 and (c) 320 nm. The closed symbols
represent the diffusion coefficient where the $R(T)$ and $L_{\phi}$
measurements have been performed. (d)-(f) Experimental data of
$R(T)$ (red solid lines) and $L_{\phi}$ (blue open symbols) as a
function of $T$; (d) $w_{\rm eff}=1130$, (e) 630 and (f) 320 nm. The
solid lines for $L_{\phi}$(T) are the best fits of
Eq.~(\ref{AAK_fitting_1D}). For the dashed-dotted lines, we have
changed the exponent of $L_{\phi}(T)$ at the low temperature part
from $-1/3$ (AAK) to $-0.29$ ($w_{\rm eff}=1130$ nm), $-0.26$
($w_{\rm eff}=630$ nm) or $-0.24$ ($w_{\rm eff}=320$ nm) in order to
get better fitting precisions down to lower temperatures. The
broken, dotted, dashed-dotted lines for $R(T)$ show the best fits of
$1/\sqrt{T}$, $\exp(T_{0}/T)$, and $\exp((T_{M}/T)^{1/2})$ laws,
respectively. The vertical dotted lines in (e) and (f) represent the
temperatures below which $L_{\phi}(T)$ and $R(T)$ deviate from the
AAK theory Eq.~(\ref{AAK_fitting_1D}) and from the activation law
Eq.~(\ref{activation_law}), respectively.}
\label{strongly_localized_regime_wires}
\end{center}
\end{figure*}

In the case of quasi-1D wires, the localization length $\xi_{\rm
loc}^{\rm 1D}$ depends on the effective width of the wires and the
diffusion coefficient as below:~\cite{gershenson_prl_97,gershenson_prb_98}
\begin{eqnarray}
\xi_{\rm loc}^{\rm 1D} = \frac{k_{F}l_{e}}{\pi}w_{\rm eff}=\frac{4m^{*}}{h}w_{\rm eff}D.
\label{localization_length_1D}
\end{eqnarray}
Since $L_{\phi}$ varies proportionally to $D^{2/3}$ in the diffusive
regime, $L_{\phi}$ can be fine tuned such that it becomes close to
$\xi_{\rm loc}^{\rm 1D}$. For the case of our wires this should
occur in the diffusion coefficient range from $D=30$ to 300
cm$^{2}/$s for $T=25$ mK.
This is shown in
Figs.~\ref{strongly_localized_regime_wires}(a)-\ref{strongly_localized_regime_wires}(c),
where we plot the theoretical localization length $\xi_{\rm
loc}^{\rm 1D}$ as well as the expected phase coherence length
$L_{\phi}^{\rm AAK}$ at our lowest temperature $T=25$ mK as a
function of $D$ for three different widths of the wires.

In
Figs.~\ref{strongly_localized_regime_wires}(d)-\ref{strongly_localized_regime_wires}(f),
we show the temperature dependence of measured $R(T)$ and
$L_{\phi}$. Here, $L_{\phi}$ has been obtained again by fitting the
magnetoconductance to the WL
theory.~\cite{note_strongly_localized_regime} Above 200 mK, the
resistance of the wires still follows a $1/\sqrt{T}$ law which is
attributable to the AA correction in the diffusive regime. Below
this temperature, the resistance deviates from the $1/\sqrt{T}$ law
and diverges exponentially. On the other hand, the phase coherence
length $L_{\phi}$ follows again a power law, but with an exponent
smaller compared to the diffusive regime [see dashed-dotted line for
$L_{\phi} (T)$ in Fig.~\ref{strongly_localized_regime_wires}]. The
qualitative behavior is indeed similar to the 2D case.

The exponential divergence of $R(T)$ can be fitted to different
exponential laws, like the simple activation
law:~\cite{thouless_prl_77}
\begin{eqnarray}
R(T) \propto \exp(T_{0}/T),
\label{activation_law}
\end{eqnarray}
or the 1D variable range hopping law:
\begin{eqnarray}
R(T) \propto \exp\left\{ (T_{M}/T)^{1/2} \right\}.
\label{VRH_law}
\end{eqnarray}
For instance, $R(T)$ for $w_{\rm eff}=1130$ nm wide wires nicely
follows Eq.~(\ref{activation_law}) down to the lowest
temperature,~\cite{note_electron_temperature_SL} whereas the
variable range hopping does not give satisfactory results. On the
other hand, for $w_{\rm eff}=630$ and 320 nm wide wires, it can also
be fitted by the activation law Eq.~(\ref{activation_law}) down to
$\sim 150$ mK, but the 1D variable range hopping law
Eq.~(\ref{VRH_law}) gives better fitting precisions down to lower
temperatures. The two fitting parameters $T_{0}$ in
Eq.~(\ref{activation_law}) and $T_{M}$ in Eq.~(\ref{VRH_law}) are
listed in Table~\ref{table3}.

\begin{table}[bmp]
\caption{Fitting parameters of the activation and 1D variable range hopping laws.}
\label{table3}
\begin{ruledtabular}
\begin{tabular}{ccccc}
$w_{\rm eff}$ (nm)&$D$ (cm$^{2}/$s)&$T_{0}$ (mK)&$T_{M}$ (mK) &$T_{\xi}$ (mK)  \\ \hline
1130&71&25& &9\\
630&170&45&30&12\\
320&290&51&39&28\\
\end{tabular}
\end{ruledtabular}
\end{table}

Similar behavior of $L_{\phi} (T)$ and exponential divergence of
resistance in quasi-1D conductors have already been reported by
Gershenson and
co-workers.~\cite{gershenson_prl_97,gershenson_prl_98} They claim
that (i) the exponential divergence of resistance is due to the
activation law, (ii) the crossover temperature $T_{\xi}$ where
$L_{\phi}^{\rm AAK}(T_{\xi})=\xi_{\rm loc}^{\rm 1D}$ is close to
$T_{0}$, and (iii) $L_{\phi}$ deviates (saturates) at certain
temperature ($T_{\rm dev}$) as the temperature approaches $T_{0}$.
These observations are \textit{qualitatively} consistent with our
experimental data. However, we observe clear \textit{quantitative}
disagreement among the three different temperatures $T_{\xi}$,
$T_{\rm dev}$ and $T_{0}$ which are more or less similar in
Refs.~\onlinecite{gershenson_prl_97} and~\onlinecite{gershenson_prl_98}. 
It is therefore
highly desirable to investigate theoretically the detailed
mechanisms of $L_{\phi}$ and $R(T)$ in quasi-1D conductors near the
crossover point from the weakly localized to strongly localized
regime.

In this section we have confirmed that in the strongly localized
regime the phase coherence time is diverging with a power law at low
temperatures. The exponent is reduced compared to the weakly
localized regime when the system approaches the strongly localized
regime. Let us remind however that for the extraction of the
exponent we applied the WL formula in a regime where it should in
principle not be valid. On the other hand, our data seems to show
that the WL theory gives a very good description of the
magnetoconductance of quasi-1D and 2D mesoscopic conductors beyond
the weakly localized regime both in the semi-ballistic and strongly
localized regimes.

\section{Conclusions}
We have studied the disorder dependence of the phase coherence
time $\tau_{\phi}$ of quasi one-dimensional (1D) wires and
two-dimensional (2D) Hall bars made from a 2D electron gas. By
implanting locally gallium ions into the doping layer of the
heterostructure using a Focused Ion Beam microscope, we have been
able to change the electronic diffusion coefficient $D$ over three
orders of magnitude. This allowed to explore various physical
regimes, namely the semi-ballistic, the weakly localized and the
strongly localized regimes. In the weakly localized regime, the
temperature as well as the diffusion coefficient dependence of the
phase coherence time is in extremely good agreement with the
``standard model" of decoherence proposed by Altshuler, Aronov and
Khmelnitsky (AAK). In particular, for quasi-1D wires, the diffusion
coefficient dependence of the phase coherence time follows a
$D^{1/3}$ power law, while the temperature dependence follows a
$T^{-2/3}$ power law. Similar observations have been found for the
2D system: the phase coherence time $\tau_{\phi}$ follows a $T^{-1}$
law as expected within the AAK theory. We do not see any sign of
saturation of the phase coherence time down to a temperatures of 25
mK. In the semi-ballistic regime where the elastic mean free path is
larger than the width of the wires, we have found a new regime where
$\tau_{\phi}$ is independent of the diffusion coefficient. In this
regime, the temperature dependence of $\tau_{\phi}$ is identical to
that of the one observed in the weakly localized regime. In the
strongly localized regime, where the resistance diverges
exponentially with decreasing temperature, we still observe a
diverging phase coherence time, however the exponent of the power
law is decreased compared to the weakly localized regime.

\begin{acknowledgments}
We acknowledge helpful discussions with G. Montambaux, C. Texier, J.
Meyer, S. Kettemann, A. D. Zaikin, S. Florens, R. Whitney, D.
Carpentier, J. V. Delft, O. Yevtushenko and C. Strunk. Y.~N.
acknowledges financial support from the \textquotedblleft JSPS
Research program for Young Scientists\textquotedblright. This work
has been supported by the European Commission FP6 NMP-3 project
505457-1 \textquotedblleft Ultra 1D\textquotedblright~and the
\textsl{Agence Nationale de la Recherche} under the grant ANR PNano
\textquotedblleft QuSpin\textquotedblright.
\end{acknowledgments}

\newpage

\begin{thebibliography}{00}
\bibitem[*]{yasu}{Present address: Institute for Solid State Physics, University of Tokyo, Japan.}
\bibitem[$\dagger$]{mail}{\texttt{bsm@listes.grenoble.cnrs.fr}}
\bibitem{AAK_82}B. L. Altshuler, A. G. Aronov and D. E. Khmelnitsky, J. Phys. C {\bf 15}, 7367 (1982).
\bibitem{mohanty_prl_97} P. Mohanty, E. M. Q. Jariwala, and R. A. Webb, Phys. Rev. Lett. {\bf 78}, 3366 (1997).
\bibitem{GZ_prl_98} D. S. Golubev and A. D. Zaikin, Phys. Rev. Lett. {\bf 81}, 1074 (1998).
\bibitem{GZ_prb_06} D. S. Golubev and A. D. Zaikin, Phys. Rev. B {\bf 74}, 245329 (2006).
\bibitem{imry_epl_99} Y. Imry, H. Fukuyama and P. Schwab, Europhys. Lett. {\bf 47}, 608 (1999).
\bibitem{zawa_prl_99} A. Zawadowski, J. von Delft, and D. C. Ralph, Phys. Rev. Lett. {\bf 83}, 2632 (1999).
\bibitem{birge+pierre_prl_02} F. Pierre and N. O. Birge, Phys. Rev. Lett. {\bf 89}, 206804 (2002).
\bibitem{schopfer_prl_03} F. Schopfer, C. B\"{a}uerle, W. Rabaud, and L. Saminadayar, Phys. Rev. Lett. {\bf 90}, 056801 (2003).
\bibitem{pierre_prb_03} F. Pierre, A. B. Gougam, A. Anthore, H. Pothier, D. Esteve, and N. O. Birge, Phys. Rev. B {\bf 68}, 085413 (2003).
\bibitem{bauerle_prl_05} C. B\"{a}uerle, F. Mallet, F. Schopfer, D. Mailly, G. Eska, and L. Saminadayar, Phys. Rev. Lett. {\bf 95}, 266805 (2005).
\bibitem{birge_prl_06} G. M. Alzoubi and N. O. Birge, Phys. Rev. Lett. {\bf 97}, 226803 (2006).
\bibitem{mallet_prl_06} F. Mallet, J. Ericsson, D. Mailly, S. \"{U}nl\"{u}bayir, D. Reuter, A. Melnikov, A. D. Wieck, T. Micklitz, A. Rosch, T. A. Costi, L. Saminadayar, and C. B\"{a}uerle, Phys. Rev. Lett. {\bf 97}, 226804 (2006).
\bibitem{sami_physicaE_07} L. Saminadayar, P. Mohanty, R. A. Webb, P. Degiovanni, and C. B\"{a}uerle, Physica E {\bf 40}, 12 (2007).
\bibitem{capron_prb_08} T. Capron, Y. Niimi, Y. Baines, D. Mailly, F.-Y. Lo, A. D. Wieck, A. Melnikov, L. Saminadayar, and C. B\"{a}uerle, Phys. Rev. B \textbf{77}, 033102 (2008).
\bibitem{glazman_03} M. G. Vavilov, L. I. Glazman, and A. I. Larkin, Phys. Rev. B {\bf 68}, 075119 (2003).
\bibitem{zarand_prl_04} G. Zar\'{a}nd, L. Borda, J. von Delft, and N. Andrei, Phys. Rev. Lett. {\bf 93}, 107204 (2004).
\bibitem{Borda_07} L. Borda, L. Fritz, N. Andrei, and G. Zar\'{a}nd, Phys. Rev. B {\bf 75}, 235112 (2007).
\bibitem{rosch_prl_06} T. Micklitz, A. Altland, T. A. Costi, and A. Rosch, Phys. Rev. Lett. {\bf 96}, 226601 (2006).
\bibitem{Micklitz_07} T. Micklitz, T. A. Costi, and A. Rosch, Phys. Rev. B {\bf 75}, 054406 (2007).
\bibitem{Costi_prl_08} T. A. Costi, L. Bergqvist, A. Weichselbaum, J. von Delft, T. Micklitz, A. Rosch, P. Mavropoulos, P. H. Dederichs, F. Mallet, L. Saminadayar, and C. B\"{a}uerle, Phys. Rev. Lett. {\bf 102}, 056802 (2009).
\bibitem{gershenson_prl_06} J. Wei, S. Pereverzev, and M. E. Gershenson, Phys. Rev. Lett. {\bf 96}, 086801 (2006).
\bibitem{Lin_2001} J. J. Lin and L. Y. Kao, J. Phys.: Condens. Matter {\bf 13}, L119 (2001).
\bibitem{Noguchi_JAP_96} M. Noguchi, T. Ikoma, T. Odagiri, H. Sakakibara, and S. N. Wang, J. Appl. Phys. {\bf 80}, 5138 (1996).
\bibitem{niimi}
Y. Niimi, Y. Baines, T. Capron, D. Mailly, F.-Y. Lo, A. D. Wieck, T. Meunier, L. Saminadayar, and C. B\"{a}uerle, Phys. Rev. Lett. {\bf 102}, 226801 (2009).
\bibitem{shallow_etching}
Y. Lee, G. Faini and D. Mailly, Phys. Rev. B {\bf 56}, 9805 (1997) and references therein.
\bibitem{fib}
J. F. Ziegler and J. P. Biersack, http://www.srim.org/.
\bibitem{note_implantation}
For this reason, we did not observe any differences on decoherence between Ga$^{+}$ and Mn$^{+}$ implantated samples with low doses ($< 10^{8}$ cm$^{-2}$), although manganese atom is a magnetic impurity.
\bibitem{wieck_pss_08} D. Diaconescu, A. Goldschmidt, D. Reuter, and A. D. Wieck, Phys. Stat. Sol. (b) {\bf 245}, 276 (2008).
\bibitem{wieck_surf_sci_90}
A. D. Wieck and K. Ploog, Surf. Sci. {\bf 229}, 252 (1990).
\bibitem{note_high_implantation_dose}
In case of high implantation doses, the implanted ion acts in GaAs predominantly as a double acceptor, which is not the case here.
\bibitem{note_D}
These temperatures have been chosen so that one can extract the
residual resistance $R_{\rm res}$ in Hall bars (i.e. minimize
$R_{\rm e-ph}(T)$ as well as $R_{\rm AA}(T)$) [see
Eq.~(\ref{residual_resistance}) in Sec. VI].
\bibitem{thermocoax}
A. Zorin, Rev. Sci. Instrum. {\bf 66}, 4296 (1994).
\bibitem{JAP_Filtres}
D. C. Glattli, P. Jacques, A. Kumar, P. Pari and L. Saminadayar, J. Appl.  Phys. \textbf{81}, 7350 (1997).
\bibitem{note_e_ph}
In noble metals, the prefactor of the temperature dependence of $1/\tau_{\rm e-e} (\propto T^{2/3})$ is the same order
as that of $1/\tau_{\rm e-ph} (\propto T^{3})$ and the crossover from $1/\tau_{\rm e-e}$ to $1/\tau_{\rm e-ph}$
can be seen at around 1 K.~\cite{sami_physicaE_07} In semiconductors, however, the former can be more than 100 times larger than the latter because of small electron density. In addition, as detailed in Sec. III B,
the decoherence time without disorder Eq.~(\ref{AAK_2D_2}) becomes dominant above 1 K.
Thus, the crossover from $1/\tau_{\rm e-e} (\propto T^{2})$ to $1/\tau_{\rm e-ph} (\propto T^{3})$ occurs at about 100 K.
This is the reason why the decoherence time due to the e-ph interactions in semiconductors can be neglected below 10 K.
\bibitem{fukuyama}
H. Fukuyama and E. Abrahams, Phys. Rev. B {\bf 27}, 5976 (1983).
\bibitem{note_crossover_T}
The crossover temperature for metals is of the order of 100 K.
\bibitem{schopfer_prl_07} F. Schopfer, F. Mallet, D. Mailly, C. Texier, G. Montambaux, C. B\"{a}uerle and L. Saminadayar, Phys. Rev. Lett. {\bf 98}, 026807 (2007).
\bibitem{texier_cmat_09} C. Texier, P. Delplace, G. Montambaux, arXiv:0907.3133 (2009).
\bibitem{birge_prb_99} D. Hoadley, P. McConville, and N. O. Birge,
Phys. Rev. B \textbf{60}, 5617 (1999).
\bibitem{mohanty_prl_03} P. Mohanty and R.A. Webb, Phys. Rev. Lett.
\textbf{91}, 066604 (2003).
\bibitem{sami_encyclopedia} L. Saminadayar, C. B\"{a}uerle, D. Mailly, \textit{in Encyclopedia of Nanoscience and Nanotechnology} ed. H. S. Nalwa, Volume 3, 267-285 (2004).
\bibitem{gilles_book} E. Akkermans and G. Montambaux, \textit{Mesoscopic physics of electrons and photons} (Cambridge University Press, Cambridge, 2007).
\bibitem{Beenakker_ssp_91} C. W. J. Beenakker and H. van Houten, Solid State Phys. {\bf 44}, 1 (1991).
\bibitem{ferrier_PRL_04}
M. Ferrier, L. Angers, A. C. H. Rowe, S. Gu\'{e}ron, H. Bouchiat, C. Texier, G. Montambaux, and D. Mailly, Phys. Rev. Lett. {\bf 93}, 246804 (2004).
\bibitem{note_fitting_L_phi}
In this fitting, we neglect the $\ln(T)$ term in Eq.~(\ref{AAK_2D_2}).
\bibitem{Renard_prb_2005}
V. T. Renard, I. V. Gornyi, O. A. Tkachenko, V. A. Tkachenko, Z. D. Kvon, E. B. Olshanetsky, A. I. Toropov and J.-C. Portal, Phys. Rev. B {\bf 72}, 075313 (2005).
\bibitem{2DEG_prb_2006}
M. Eshkol, E. Eisenberg, M. Karpovski, and A. Palevski, Phys. Rev. B {\bf 73}, 115318 (2006).
\bibitem{2DHG_prb_2004}
S. McPhail, C. E. Yasin, A. R. Hamilton, M. Y. Simmons, E. H. Linfield, M. Pepper, and D. A. Ritchie, Phys. Rev. B {\bf 70}, 245311 (2004).
\bibitem{Beenakker_prb_88}
C. W. J. Beenakker and H. van Houten, Phys. Rev. B {\bf 38}, 3232 (1988).
\bibitem{boundary_prl_89}
T. J. Thornton, M. L. Roukes, A. Scherer, and B. P. Van de Gaag, Phys. Rev. Lett. {\bf 63} 2128 (1989).
\bibitem{ando_prb_91}
H. Akera and T. Ando, Phys. Rev. B {\bf 43}, 11676 (1991).
\bibitem{boundary_calculation_66}
E. Ditlefsen and J. Lothe, Philos. Mag. {\bf 14}, 759 (1966).
\bibitem{bouchiat}
B. Reulet, H. Bouchiat and D. Mailly, Europhys. Lett. {\bf 31}, 305 (1995).
\bibitem{aleiner_prb_02}
B. N. Narozhny, G. Zala, and I. L. Aleiner, Phys. Rev. B {\bf 65}, 180202(R) (2002).
\bibitem{note_definition_D}
It is also possible to obtain a diffusion coefficient from quasi-1D wires which is 1.5 times larger than that from Hall bars. Plotting the data as a function of $D$ from the wires rather than the 2D Hall bars would simply shift all the data by a fixed value. This does not change the $D$ dependence as well as the interpretation of the data at all. Nevertherless, we have chosen to plot all the data as a function of $D$ from the 2D Hall bars because it is difficult to obtain, for example, $n_{e}$ and $l_{e}$ only from the wires. In order to define the regime (semi-ballistic or diffusive), it is necessary to know $l_{e}$. Therefore we have determined $D$ from the Hall bars.
\bibitem{note_residual_resistance}
The residual resistance $R_{\rm res}$ defined here is for 20 wires in parallel. The resistance per 1 wire is 20 times larger. When $R_{\rm res}>8$ k$\Omega$, the resistance exponentially increases with decreasing temperature, which indicates that the electron system enters the strongly localized regime as detailed in Sec. VII.
\bibitem{wittmann_jltp_87}
H.-P. Wittmann and A. Schmid, J. Low Temp. Phys. {\bf 69}, 131 (1987).
\bibitem{GZ_physica_E_07}
D. S. Golubev and A. D. Zaikin, Physica E {\bf 40}, 32 (2007) and references therein.
\bibitem{GZ_prb_99}
D. S. Golubev and A. D. Zaikin, Phys. Rev. B {\bf 59}, 9195 (1999).
\bibitem{GZ_jltp_02}
D. S. Golubev, A. D. Zaikin and G. Sch\"{o}n, J. Low Temp. Phys. {\bf 126}, 1355 (2002).
\bibitem{GZ_metallic_wire}
In metallic wires, the 3D formula of the GZ theory is applied since all geometric dimensions ($L,w$ and film thickness $t$) are larger than $l_{e}$.
\bibitem{AA_correction}
B. L. Altshuler and A. G. Aronov, in \textit{Electron-Electron Interactions in Disordered Systems}, edited by A. L. Efros and M. Pollak (North-Holland, Amsterdam, 1985).
\bibitem{note_L_T} 
One could expect a dimensional
crossover for the AA correction when the thermal length $L_T =
\sqrt{\hbar D/k_B T}$ is larger than $w_{\rm eff}$. In this case one
would expect a deviation from the 1D law at high temperatures. For a
diffusion coefficient $D$ of 1000 cm$^{2}/$s and 
an effective width $w_{\rm eff}$ of 1~$\mu$m 
the crossover temperature is of the order of 1 K. Lowering
the diffusion coefficient, the crossover temperature should in
priciple move towards lower temperatures. On the other hand, for all
diffusion coefficients investigated as well as for all wire widths,
we do not observe such a dimensional crossover at temperatures below
1 K. This suggests that the 1D-2D crossover appears only for $L_T$
much smaller than $w_{\rm eff}$.
\bibitem{gornyi_prb_04}
I. V. Gornyi and A. D. Mirlin, Phys. Rev. B {\bf 69}, 045313 (2004).
\bibitem{zaikin_prb_04} D. S. Golubev and A. D. Zaikin, Phys. Rev. B {\bf 70}, 165423 (2004).
\bibitem{mora_prb_07}
C. Mora, R. Egger, and A. Atland, Phys. Rev. B {\bf 75}, 035310
(2007).
\bibitem{faran_prb_88}
O. Faran and Z. Ovadyahu, Phys. Rev. B {\bf 38}, 5457 (1988).
\bibitem{hsu_prl_95}
S.-Y. Hsu and J. M. Valles, Phys. Rev. Lett. {\bf 74}, 2331 (1995).
\bibitem{tremblay_jp_90}
F. Tremblay, M. Pepper, R. Newbury, D. A. Ritchie, D. C. Peacock, J. E. F. Frost, G. A. C. Jones, and G. Hill, J. Phys.; Condens. Matter {\bf 2}, 7367 (1990).
\bibitem{jiang_prb_92}
H. W. Jiang, C. E. Johnson, and K. L. Wang, Phys. Rev. B {\bf 46}, 12830 (1992).\bibitem{keuls_prb_97}
F. W. Van Keuls, X. L. Hu, H. W. Jiang, and A. J. Dahm, Phys. Rev. B {\bf 56}, 1161 (1997).
\bibitem{khondaker_prb_99}
S. I. Khondaker, I. S. Shlimak, J. T. Nicholls, M. Pepper, and D. A. Ritchie, Phys. Rev. B {\bf 59}, 4580 (1999).
\bibitem{minkov_prb_02}
G. M. Minkov, O. E. Rut, A. V. Germanenko, A. A. Sherstobitov, B. N. Zvonkov, E. A. Uskova, and A. A. Birukov, Phys. Rev. B {\bf 65}, 235322 (2002).
\bibitem{minkov_prb_04}
G. M. Minkov, A. V. Germanenko, and I. V. Gornyi, Phys. Rev. B {\bf 70}, 245423 (2004).
\bibitem{minkov_prb_07}
G. M. Minkov, A. V. Germanenko, O. E. Rut, A. A. Sherstobitov, B. N. Zvonkov, Phys. Rev. B {\bf 75}, 235316 (2007).
\bibitem{peled_prl_03}
E. Peled, D. Shahar, Y. Chen, E. Diez, D. L. Sivco, and A. Y. Cho, Phys. Rev. Lett. {\bf 91}, 236802 (2003).
\bibitem{li_prl_09}
W. Li, C. L. Vicente, J. S. Xia, W. Pan, D. C. Tsui, L. N. Pfeiffer, and K. W. West, Phys. Rev. Lett. {\bf 102}, 216801 (2009).
\bibitem{efros_jp_75}
A. L. Efros and B. I. Shklovskii, J. Phys. C {\bf 8} L49 (1975).
\bibitem{vollhardt_prl_80}
D. Vollhardt and P. W\"{o}lfle, Phys. Rev. Lett. {\bf 45}, 842 (1980); Phys. Rev. B {\bf 22}, 4666 (1980).
\bibitem{mott_jp_81}
N. F. Mott and M. Kaveh, J. Phys. C {\bf 14}, L659 (1981).
\bibitem{davies_jp_82}
R. A. Davies and M. Pepper, J. Phys. C {\bf 15}, L371 (1982).
\bibitem{gogolin_ssc_93}
A. A. Gogolin and G. T. Zim\'{a}nyi, Solid State Commun. {\bf 46},
469 (1983).
\bibitem{zhao_prb_91}
H. L. Zhao, B. Z. Spivak, M. P. Gelfand, and S. Feng, Phys. Rev. B
{\bf 44}, 10760 (1991).
\bibitem{note_self_consistent_theory}
It is difficult to distinguish clearly between the variable range
hopping model and the self-consistent theory from our results. The
same analyses for the conductivity in the strongly localized regime
have been performed by Minkov \textit{et al}.~\cite{minkov_prb_02}
But they could not identify a reliable mechanism of the
conductivity, either.
\bibitem{note_L_N_2D}
Note that Eq.~(\ref{AAK_2D_1}) is valid only in the weakly localized
regime and should not be applicable in the strongly localized
regime.
\bibitem{gershenson_prl_97}
M. E. Gershenson, Y. B. Khavin, A. G. Mikhalchuk, H. M. Bozler, and A. L. Bogdanov, Phys. Rev. Lett. {\bf 79}, 725 (1997).
\bibitem{gershenson_prb_98}
Y. B. Khavin, M. E. Gershenson, and A. L. Bogdanov, Phys. Rev. B {\bf 58}, 8009 (1998).
\bibitem{note_strongly_localized_regime}
Although $L_{\phi}$ for $w_{\rm eff}=1130$ nm wide wire is smaller than $w_{\rm eff}$ above $T=100$ mK, the WL curves can still be fitted
well by the 1D WL theory Eq.~(\ref{HLN}) over the whole temperature range.
\bibitem{thouless_prl_77}
D. J. Thouless, Phys. Rev. Lett. {\bf 39}, 1167 (1977).
\bibitem{note_electron_temperature_SL} In
Figs.~\ref{strongly_localized_regime_wires}(d)-\ref{strongly_localized_regime_wires}(f),
we have already corrected the electron temperature below 60 mK by
using the AA law for another diffusive wire or Hall bar at the same
diffusion coefficient $D$.
\bibitem{gershenson_prl_98}
Y. B. Khavin, M. E. Gershenson, and A. L. Bogdanov, Phys. Rev. Lett. {\bf 81}, 1066 (1998).
\end{thebibliography}

\end{document}